\documentclass[preprint,aps,12pt,showpacs,nofootinbib,tightenlines]{revtex4}
\usepackage{amsmath}
\usepackage{amssymb}
\usepackage{epsfig}
\usepackage{graphicx}
\begin{document}

\newcommand{\beq}{\begin{eqnarray}}
\newcommand{\eeq}{\end{eqnarray}}
\newcommand{\non}{\nonumber\\ }

\newcommand{\pka}{\phi_{K}^A}
\newcommand{\pkp}{\phi_K^P}
\newcommand{\pkt}{\phi_K^T}

\newcommand{\pea}{\phi_{\eta}^A}
\newcommand{\pep}{\phi_{\eta}^P}
\newcommand{\pet}{\phi_{\eta}^T}
\newcommand{\peqa}{\phi_{\eta_q}^A}
\newcommand{\peqp}{\phi_{\eta_q}^P}
\newcommand{\peqt}{\phi_{\eta_q}^T}

\newcommand{\pesa}{\phi_{\eta_s}^A}
\newcommand{\pesp}{\phi_{\eta_s}^P}
\newcommand{\pest}{\phi_{\eta_s}^T}

\newcommand{\pepa}{\phi_{\eta'}^A}
\newcommand{\pepp}{\phi_{\eta'}^P}
\newcommand{\pept}{\phi_{\eta'}^T}

\newcommand{\pksa}{\phi_{K^*}}
\newcommand{\pksp}{\phi_{K^*}^s}
\newcommand{\pkst}{\phi_{K^*}^t}

\newcommand{\fb}{f_B }
\newcommand{\fk}{f_K }
\newcommand{\fe}{f_{\eta} }
\newcommand{\fep}{f_{\eta'} }
\newcommand{\rks}{r_{K^*} }
\newcommand{\rk}{r_K }
\newcommand{\re}{r_{\eta} }
\newcommand{\rep}{r_{\eta'} }
\newcommand{\mb}{m_B }
\newcommand{\mw}{m_W }

\newcommand{\kks}{K^{(*)}}
\newcommand{\etar}{\eta^\prime }
\newcommand{\acp}{{\cal A}_{CP}}
\newcommand{\etap}{\eta^{(\prime)} }
\newcommand{\pb}{\phi_B}
\newcommand{\pks}{\phi_{K^*}}

\newcommand{\xeba}{\bar{x}_2}
\newcommand{\xsba}{\bar{x}_3}
\newcommand{\res}{r_{\eta_s}}
\newcommand{\red}{r_{\eta_q}}
\newcommand{\peas}{\phi^A_{\eta_s}}
\newcommand{\peps}{\phi^P_{\eta_s}}
\newcommand{\pets}{\phi^T_{\eta_s}}
\newcommand{\pead}{\phi^A_{\eta_q}}
\newcommand{\pepd}{\phi^P_{\eta_q}}
\newcommand{\petd}{\phi^T_{\eta_q}}

\newcommand{\pvsl}{ p \hspace{-2.0truemm}/_{K^*} }
\newcommand{\esl}{ \epsilon \hspace{-2.1truemm}/ }
\newcommand{\psl}{ P \hspace{-2.4truemm}/ }
\newcommand{\nsl}{ n \hspace{-2.2truemm}/ }
\newcommand{\vsl}{ v \hspace{-2.2truemm}/ }
\newcommand{\epsl}{\epsilon \hspace{-1.8truemm}/\,  }
\newcommand{\bfkk}{{\bf k} }
\newcommand{\calm}{ {\cal M} }
\newcommand{\calh}{ {\cal H} }

\def \epjc{ Eur. Phys. J. C }
\def \jpg{  J. Phys. G }
\def \npb{  Nucl. Phys. B }
\def \plb{  Phys. Lett. B }
\def \pr{  Phys. Rep. }
\def \prd{  Phys. Rev. D }
\def \prl{  Phys. Rev. Lett.  }
\def \zpc{  Z. Phys. C  }
\def \jhep{ J. High Energy Phys.  }

\title{Branching ratios and CP asymmetries of $B \to K \eta^{(\prime)}$ decays
in the  pQCD approach }

\author{Zhen-Jun Xiao \footnote{Electronic address: xiaozhenjun@njnu.edu.cn},
Zhi-Qing Zhang, Xin Liu, and Li-Bo Guo
\footnote{Electronic address: guolibo@njnu.edu.cn} }
\affiliation{Department of Physics and Institute of
Theoretical Physics,
Nanjing Normal University, Nanjing, Jiangsu 210097, P.R.China }
\date{\today}
\begin{abstract}
We calculate the branching ratios and CP violating asymmetries
of the four $B \to K \etap$ decays in the perturbative QCD (pQCD)
factorization approach. Besides the full leading order contributions, the partial
next-to-leading order (NLO) contributions from the QCD vertex corrections, the quark loops,
and the chromo-magnetic penguins are also taken into account.
The NLO pQCD predictions for the CP-averaged branching ratios are
$Br(B^+ \to K^+ \eta) \approx 3.2 \times 10^{-6}$,
$Br(B^\pm \to K^\pm \etar) \approx 51.0 \times 10^{-6}$,
$Br(B^0 \to K^0 \eta) \approx 2.1 \times 10^{-6}$,
and $Br(B^0 \to K^0 \etar) \approx 50.3 \times 10^{-6}$.
The NLO contributions can provide a $70\%$ enhancement to the LO $Br(B \to K \etar)$, but
a $30\%$ reduction to the LO $Br(B \to K \eta)$, which play the key role in
understanding the observed pattern of branching ratios.
The NLO pQCD predictions for the CP-violating asymmetries, such as
$\acp^{dir} (K^0_S \etar) \sim 2.3\% $ and $\acp^{mix}(K^0_S \etar)\sim 63\%$,
agree very well with currently available data.
This means that the deviation
$\Delta S=\acp^{mix}(K^0_S \etar) - \sin{2\beta}$ in pQCD approach
is also very small.
\end{abstract}

\vspace{1cm}


\maketitle

\section{Introduction}

The $B\to K\etap$ decays are very interesting
two-body charmless hadronic B meson decays. In 1997, CLEO collaboration  firstly
reported unexpectedly large branching ratios for $B \to K \etar$ decays \cite{cleo97}.
Eleven years later, three of the four $B\to K \etap$ decays have been measured with
high  precision. The world averages as given by HFAG \cite{hfag} are the
following (in unit of $10^{-6}$)
\beq
Br(B^\pm \to K^\pm \eta)&=& 2.7 \pm 0.3, \non
Br(B^\pm \to K^\pm \etar)&=& 70.2\pm 2.5, \non
Br(B^0 \to K^0 \eta)&<& 1.9, \non
Br(B^0 \to K^0 \etar)&=&64.9\pm 3.1. \label{eq:br-k}
\eeq
From above data one can see that: (a) the measured $Br(B \to K \etar)$ are much
larger than the early standard model (SM) expectations, i.e.,  the so-called $k\etar$-puzzle;
and (b) the large disparity between the branching ratios for $B \to K \etar$ and
$B \to K \eta$ decays: $Br(B\to K \etar) \gg Br(B \to K \eta)$.

Besides the branching ratios, the CP violating asymmetries for $B^\pm \to K^\pm \etap$
and $B^0 \to K^0 \etap$ decays have been measured very recently \cite{hfag,acpketap}:
\beq
\acp^{dir}(B^\pm \to K^\pm \eta)&=& -0.27 \pm 0.09, \non
\acp^{dir}(B^\pm \to K^\pm \etar)&=& 0.016 \pm 0.019, \\
\acp^{dir}(B^0 \to K^0 \etar)&=& 0.09 \pm 0.06, \non
\acp^{mix}(B^0 \to K^0 \etar)&=& 0.61 \pm 0.07,
\label{eq:acp01}
\eeq
It may be noted that the average of the measured $\acp^{mix}(B^0 \to K^0 \etar)$ is
now more than $8 \sigma$
away from zero, so that CP violation in this decay is well established; while
$\acp^{dir}(B^0 \to K^0 \etar)$ is not conflict with zero as expected in the SM.
The data for $\acp^{dir}(B^\pm \to K^\pm \etap)$ have less precision, but are consistent
with general expectations.

The measurements of time-dependent CP asymmetries in $B^0$ meson decays, such as
$B^0 \to J/\Psi K^0$ via $b \to c \bar{c} s $  ``tree" transition and
$B^0 \to K^0 \etar$ via $b \to s q\bar{q}$ penguin transition, have provided crucial
tests of the mechanism of CP violation in the SM. Within the SM the mixing induced
CP violating asymmetry $\acp^{mix}(B^0 \to K^0 \etar)=-\eta_f S_f $ should be comparable
with $\sin 2\beta=0.685$ obtained from the tree dominated $B^0 \to J/\Psi K^0$ decay,
this point has been confirmed by the data in Eq.~(\ref{eq:acp01}).

In the SM  the decay $B \to K \etap$ is believed to proceed dominantly through
gluonic  penguin processes\cite{lipkin91,grossman} and has been evaluated
by employing various methods
\cite{as97,chao97,ali98,du426,yy01,ahmady98,kagan,xiao99,bn651,npb675,kou02}.
Although great progress have been made during the past decade, but the predictions for
$Br(B \to K \etar)$ from both the QCD  factorization (QCDF) approach \cite{bn651,bbns99}
and the perturbative QCD (pQCD) approach \cite{kou02,li2003}
in the Feldmann-Kroll-Stech (FKS) mixing
scheme of $\eta-\etar$ system \cite{fks98,jhep0506} are smaller than the data.

For the pattern of branching ratios in Eq.~(\ref{eq:br-k}), many possible
solutions have been proposed. These include, for example,
\begin{itemize}
\item[]{(a)}
Conventional $b\to s q\bar{q}$ with constructive (destructive)
interference between the $u\bar{u}, d\bar{d}$ and $s\bar{s}$ components of $\etar$ ($\eta$)
\cite{lipkin91};

\item[]{(b)}
Large intrinsic charm content of $\eta^{\prime}$ through the
chain  $b\to s c \bar c \to s \eta^{\prime} $ \cite{chao97} or through
$b\to s c \bar c \to s g^*g^*\to s (\eta,\etar)$ due to the QCD anomaly \cite{ali98};

\item[]{(c)}
The spectator hard-scattering  mechanism through the anomalous
coupling of $gg \to \eta^\prime$ \cite{du426,yy01,ahmady98};

\item[]{(d)}
A significant flavor-singlet contribution \cite{yy01,bn651};

\item[]{(e)}
A strong penguin $b\to sg$ enhanced by new physics \cite{kagan,xiao99}.

\end{itemize}
But the data of branching ratio in Eq.~(\ref{eq:br-k}) are still not completely understood.
For the CP violation of $B \to K \etap$ decays, the theoretical studies is still under way.

In Ref.~\cite{kou02}, the authors calculated the branching ratios of $B\to K \etap$ decays
by employing the pQCD  approach at leading order. They considered the large corrections from
$SU(3)$ flavor symmetry breaking as well as the possible gluonic component of $\etar$ meson, but
their prediction for $Br(B^0 \to K^0 \etar)$ ( $Br(B^0 \to K^0 \eta)$ ) is much
smaller ( larger) than the measured value.

A sizable gluonic content in $\etar$ meson may provide a large enhancement to the decay rate of
$B\to K \etar$. In Ref.~\cite{li0609}, the authors examined the possible
gluonic contribution to the $B\to \etar$ transition form factor and found that
such contribution is constructive with those from quark-content of $\etar$, but
numerically very small and can be neglected safely.
This point has also been confirmed by the QCD sum-rule analysis\cite{bj0706}

In the  quark-flavor mixing scheme, the physical $\eta$ and $\etar$ meson
are linear combinations of flavor state $\eta_q=(u\bar{u}+d\bar{d})/\sqrt{2}$
and $\eta_s=s\bar{s}$ with the "mass" of $m_{qq}$  and $m_{ss}$ respectively.
In Ref.~\cite{acg07}, the effect of a large chiral scale $m_0^q=m_{qq}^2/(2m_q)$
with $q =(u,d)$ for the meson $\eta_q$ has been evaluated although we do not know
which mechanism is responsible to achieve a large value of $m_{qq}$.
When one uses $m_{qq}=0.22$ GeV \cite{acg07}
instead of its generally accepted value of $m_{qq}=0.11$ GeV, a larger $B\to K \eta_q$
decay amplitude can be obtained. Consequently, the LO pQCD predictions
for $Br(B \to K \etar)$ become consistent with the data.

In Ref.~\cite{hsu07}, the authors examined the possible way to increase the value
of $m_{qq}$. They
found that few-percent violation of Okubo-Zweig-Iizuka (OZI) rule can enhance $m_{qq}$
few times, which then leads to the  consistency of the LO predictions with the data
for $B \to K \etap$ decays.

Besides the possible mechanisms mentioned above, we here consider a new and natural
solution: the effects of the next-to-leading order (NLO) contributions in the pQCD approach.
As shown in Ref.~\cite{nlo05}, the NLO contributions to $B \to K \pi$ decays
can play the key rule to explain the so-called $``K\pi"-$ puzzle. We expect here
the NLO contributions could help us to resolve the $``K\etar"-$puzzle.

For the CP asymmetries of $B^0 \to K^0 \etar$, the deviation $\Delta S_f = -\eta_f S_f -\sin2\beta $
has been estimated, for example,  in the QCDF approach \cite{npb675,qcdf1} and the soft collinear
effective theory\cite{scet06}. The resultant bound is $|\Delta S_f| \lesssim 0.05$.
Since the source of the CP violation in the pQCD approach is very different from those in the
QCDF/SCET approach, we here try to calculate the CP asymmetries of $B \to K \etap$ decays
by employing the pQCD approach at LO and NLO level, to check if we can accommodate the data
of CP asymmetries.

In this paper  we will calculate the next-to-leading
order contributions to the branching ratios and CP violating asymmetries of the
four $B \to K \etap$  decays.
We firstly calculate the decay amplitudes
of the $B\to K \etap$ decays by employing the pQCD factorization
approach at the leading order (LO),
as have been done in previous studies for other two-body charmless B meson decays
\cite{kls01,luy01,xiao06,xiao07}.
And then we evaluate the NLO contributions to these decays.

The NLO contributions considered here include: QCD vertex corrections, the quark-loops
and the chromo-magnetic penguins. We wish that they are the major part of the full
NLO contributions in pQCD approach \cite{nlo05}.
Of course, remaining NLO contributions in pQCD approach, such as those from
factorizable emission diagrams, hard-spectator and annihilation diagrams,
should be calculated as soon as possible.

This paper is organized as follows. In Sec.II, we
give a brief review about the pQCD factorization approach.
In Sec.~III, we calculate analytically the relevant Feynman diagrams and
present the various decay amplitudes for the studied decay modes in
leading-order.
In Sec.~IV, the NLO contributions from the vertex corrections, the
quark loops and the chromo-magnetic penguin amplitudes are evaluated.
We calculate and show the pQCD predictions for the branching ratios and  CP violating
asymmetries of $B \to K\etap$ decays in Sec.~V.
The summary and some discussions are included in the final section.


\section{ Theoretical framework}\label{sec:f-work}

\subsection{ Theoretical framework}\label{sec:2-1}

In the pQCD approach, the decay amplitude is separated into soft ($\Phi_{M_i}$), hard
( $H(k_i,t)$ ), and harder( $C(M_W)$ ) dynamics characterized by different energy
scales $( \Lambda_{QCD}, t, m_b, M_W )$ \cite{li2003}.
The decay amplitude ${\cal A}(B \to M_2 M_3)$ can be written conceptually as the convolution,
\beq
{\cal A}(B \to M_2 M_3)\sim \int\!\! d^4k_1 d^4k_2 d^4k_3\ \mathrm{Tr}
\left [ C(t) \Phi_B(k_1) \Phi_{M_2}(k_2) \Phi_{M_3}(k_3)
H(k_1,k_2,k_3, t) \right ],
\label{eq:con1}
\eeq
where $k_i$'s are momenta of light quarks included in each meson, and $\mathrm{Tr}$
denotes the trace over Dirac and color indices. $C(t)$ is the Wilson
coefficient evaluated at scale $t$.
In the above convolution, the Wilson coefficient $C(t)$ includes the harder dynamics at scale
higher than $M_B$ and describes the
evolution of local $4$-Fermi operators from $m_W$ ( the $W$ boson
mass) down to $ t \sim \mathcal{O}(\sqrt{\bar{\Lambda} M_B})$ scale,
where $\bar{\Lambda}\equiv M_B -m_b$.  The function
$H(k_1,k_2,k_3,t)$ describes the four quark operator and the
spectator quark connected by  a hard gluon whose $q^2$ is in the order
of $\bar{\Lambda} M_B$, and includes the
$\mathcal{O}(\sqrt{\bar{\Lambda} M_B})$ hard dynamics. Therefore,
this hard kernel $H$ can be perturbatively calculated. The function
$\Phi_{M_i}$ is the wave function which describes hadronization of the
quark and anti-quark in the meson $M_i$.
While the hard kernel $H$ depends on the processes considered,
the wave function $\Phi_{M_i}$ is independent of the specific processes.
Using the wave functions determined from other well measured processes, one can make
quantitative predictions here.

Since the b quark inside the B meson is rather heavy, we consider the $B$ meson at rest
for simplicity. It is then  convenient to use light-cone coordinate $(p^+,
p^-, {\bf p}_{\rm T})$ to describe the meson's momenta:
$p^\pm = \frac{1}{\sqrt{2}} (p^0 \pm p^3)$ and ${\bf p}_{\rm T} = (p^1, p^2)$ .
Using the light-cone coordinates the $B$ meson momentum $P_B$ and the two
final state meson's momenta $P_2$ and $P_3$ (for $M_2$ and $M_3$ respectively)
can be written as
\beq
P_B = \frac{M_B}{\sqrt{2}} (1,1,{\bf 0}_{\rm T}), \quad
P_2 = \frac{M_B}{\sqrt{2}}(1-r_3^2,r^2_2,{\bf 0}_{\rm T}), \quad
P_3 = \frac{M_B}{\sqrt{2}} (r_3^2,1-r^2_2,{\bf 0}_{\rm T}),
\eeq
where $r_i=m_i/M_B$. $m_2$ and $m_3$ are the mass of the two final state mesons.
For the case of $B \to PP$ decays, $r_2$ and $r_3$ are small and
could be neglected safely.

Putting the anti-quark momenta in $B$, $M_2$ and $M_3$ meson as $k_1$, $k_2$, and $k_3$,
respectively, we can choose
\beq
k_1 = (x_1 P_1^+,0,{\bf k}_{\rm 1T}), \quad
k_2 = (x_2 P_2^+,0,{\bf k}_{\rm 2T}), \quad
k_3 = (0, x_3 P_3^-,{\bf k}_{\rm 3T}).
\eeq
Then, the integration over $k_1^-$, $k_2^-$, and $k_3^+$ in
eq.(\ref{eq:con1}) will lead to
\beq
{\cal A}(B \to P V ) &\sim
&\int\!\! d x_1 d x_2 d x_3 b_1 d b_1 b_2 d b_2 b_3 d b_3 \non &&
\cdot \mathrm{Tr} \left [ C(t) \Phi_B(x_1,b_1) \Phi_{M_2}(x_2,b_2)
\Phi_{M_3}(x_3, b_3) H(x_i, b_i, t) S_t(x_i)\, e^{-S(t)} \right ],
\quad \label{eq:a2}
\eeq
where $b_i$ is the conjugate space
coordinate of $k_{iT}$. The large logarithms ($\ln m_W/t$) coming
from QCD radiative corrections to four quark operators are included
in the Wilson coefficients $C(t)$. The large double logarithms
($\ln^2 x_i$) on the longitudinal direction are summed by the
threshold resummation, and they lead to $S_t(x_i)$ which
smears the end-point singularities on $x_i$. The last term,
$e^{-S(t)}$, is the Sudakov form factor which suppresses the soft
dynamics effectively \cite{li2003}.

\subsection{ Effective Hamiltonian and Wilson coefficients}\label{sec:2-1b}

For the studied $B \to K \etap$ decays, the weak effective Hamiltonian $H_{eff}$
for $b \to s$ transition can be written as \cite{buras96}
\beq
\label{eq:heff}
{\cal H}_{eff} = \frac{G_{F}}
{\sqrt{2}} \, \sum_{q=u,c}V_{qb} V_{qs}^*\left\{  \left [ C_1(\mu)
O_1^q(\mu) + C_2(\mu) O_2^q(\mu) \right ]
+ \sum_{i=3}^{10} C_{i}(\mu) \;O_i(\mu) \right\} \; .
\eeq
where $G_{F}=1.166 39\times 10^{-5} GeV^{-2}$ is the Fermi constant,
and $V_{ij}$ is the CKM matrix element, $C_i(\mu)$ are the Wilson coefficients evaluated
at the renormalization scale $\mu$ and $O_i(\mu)$ are the four-fermion operators.
For the case of $b \to d $ transition, simply makes a replacement of $s$ by
$d$ in Eq.~(\ref{eq:heff}) and in the
expressions of $O_i(\mu)$ operators, which can be found easily for example in
Refs.\cite{xiao06,xiao07,buras96}.

In PQCD approach, the energy scale $``t"$ is chosen as the largest energy scale in
the hard kernel $H(x_i,b_i,t)$ of a given Feynman diagram, in order to
suppress the higher order corrections and improve the reliability of the perturbative
calculation.
Here, the scale $``t"$ may be larger or smaller than the $m_b$ scale.
In the range of $ t < m_b $ or $t \geq m_b$, the number of active quarks is $N_f=4$ or
$N_f=5$, respectively.
For the Wilson coefficients $C_i(\mu)$ and their renormalization group (RG) running,
they are known at NLO level currently \cite{buras96}.
The explicit expressions of the LO and NLO $C_i(\mw)$ can be found easily, for example, in
Refs.~\cite{luy01,buras96}.

When the pQCD approach at leading-order are employed, the leading order Wilson
coefficients $C_i(m_W)$, the leading order RG evolution matrix $U(t,m)^{(0)}$ from
the high scale $m$ down to $t < m$ ( for details see Eq.~(3.94) in Ref.~\cite{buras96}),
and the leading order $\alpha_s(t)$ are used:
\beq
\alpha_s(t)=\frac{4\pi}{ \beta_0 \ln \left [ t^2/ \Lambda_{QCD}^2\right]},
\eeq
where $\beta_0 = (33- 2 N_f)/3$, $\Lambda_{QCD}^{(5)}=0.225 GeV$ and
$\Lambda_{QCD}^{(4)}=0.287$ GeV.

When the NLO contributions are taken into account, however,
the NLO Wilson coefficients $C_i(m_W)$, the NLO RG evolution matrix $U(t,m,\alpha)$
( for details see Eq.~(7.22) in Ref.~\cite{buras96}) and the $\alpha_s(t)$ at two-loop level
are used:
\beq
\alpha_s(t)=\frac{4\pi}{ \beta_0 \ln \left [ t^2/ \Lambda_{QCD}^2\right]}
\cdot \left \{ 1- \frac{\beta_1}{\beta_0^2 } \cdot
\frac{ \ln\left [ \ln\left [ t^2/\Lambda_{QCD}^2  \right]\right]}{
\ln\left [ t^2/\Lambda_{QCD}^2\right]} \right \},
\label{eq:asnlo}
\eeq
where $\beta_0 = (33- 2 N_f)/3$, $\beta_1 = (306-38 N_f)/3$, $\Lambda_{QCD}^{(5)}=0.225$ GeV and
$\Lambda_{QCD}^{(4)}=0.326$ GeV.

From the general knowledge, the hard scale $t$ must be much larger than
$\Lambda_{QCD}\approx 0.2$ GeV in order to guarantee the reliability of
perturbative calculations. In previous calculations based on the pQCD approach
$\mu_0=0.5$ GeV is chosen as the lower cut-off of the scale $t$.
In our opinion, it is indeed too low, because it may be conceptually incorrect to evaluate
the Wilson coefficients at scales down to $0.5$ GeV ~\cite{beneke07}.
The explicit numerical checks as done in Ref.~\cite{xiao08a} also show that
(a) the Wilson coefficient $C_1(0.5)$ is close to $-1$ and clearly too large in size!
(b) the values of the Wilson coefficients $C_{3,4,5,6}(\mu)$ at $\mu=0.5$ GeV are
about four to seven times larger than those at $\mu=1.0$ GeV;
and (c) the $\mu_0-$dependence of all Wilson coefficients become relatively weak for
$\mu_0\geq 1.0$ GeV.
We therefore believe that it is reasonable to choose $\mu_0=1.0$ GeV
as the lower cut-off of the hard scale $t$, which is also close to the hard-collinear scale
$\sqrt{\bar{\Lambda}m_B} \sim 1.3$ GeV in SCET.
In the numerical integrations we will
fix the values $C_{i}(t)$ at $C_{i}(1.0)$  whenever the scale $t$ runs below the
scale $\mu_0=1.0$ GeV \cite{xiao08a,xiao08b}.

\subsection{ Wave functions}\label{sec:2-3}

Since the b-quark is much heavier than the up or down quark, the $B$ meson is
treated as a very good heavy-light system.
Although there are in general two Lorentz structures in the B meson distribution
amplitudes, they obey to the following normalization conditions
\beq
\int\frac{d^4 k_1}{(2\pi)^4}\phi_B({\bf k_1})
=\frac{f_B}{2\sqrt{2N_c}}, ~~~\int \frac{d^4
k_1}{(2\pi)^4}\bar{\phi}_B({\bf k_1})=0.
\eeq
However, it can be argued that the contribution of $\bar{\phi}_B$ is numerically small
~\cite{luyang}, thus its contribution can be numerically
neglected. In this approximation, we only consider the
contribution of Lorentz structure
\beq
\Phi_B= \frac{1}{\sqrt{2N_c}}
(\psl_B +m_B) \gamma_5 \phi_B ({\bf k_1}), \label{bmeson}
\eeq
with
\beq
\phi_B(x,b)&=& N_B x^2(1-x)^2 \mathrm{exp} \left
 [ -\frac{M_B^2\ x^2}{2 \omega_{b}^2} -\frac{1}{2} (\omega_{b} b)^2\right],
 \label{phib}
\eeq
where $\omega_{b}$ is a free parameter and we take
$\omega_{b}=0.4\pm 0.04$ GeV in numerical calculations, and
$N_B=101.445$ is the normalization factor for $\omega_{b}=0.4$.

The Kaon mesons are  treated as a light-light system.
The wave function of $K$ meson is defined as \cite{ball98}
\beq
\Phi_{K}(P,x,\zeta)\equiv \frac{1}{\sqrt{2N_C}}\gamma_5 \left [ \psl
\phi_{K}^{A}(x)+m_0^{K} \phi_{K}^{P}(x)+\zeta m_0^{K} (\vsl \nsl -
v\cdot n)\phi_{K}^{T}(x)\right ],
\eeq
where $P$ and $x$ are the momentum and the momentum fraction of $K$, respectively. The
parameter $\zeta$ is either $+1$ or $-1$ depending on the assignment
of the momentum fraction $x$.

For $\etap$ meson, the wave function for $\eta_q$ components of $\etar$ meson
are given as
\beq
\Phi_{\eta_q}(P,x,\zeta)\equiv \frac{1}{\sqrt{2N_C}} \gamma_5 \left [ \psl
\phi_{\eta_q}^{A}(x)+m_0^q \phi_{\eta_q}^{P}(x)+\zeta m_0^q
( \vsl \nsl - v\cdot n)\phi_{\eta_q}^{T}(x) \right ],
\eeq
where $P$ and $x$ are the momentum and the momentum fraction of $\eta_q$, respectively.
We assumed here that the wave
function of $\eta_q$ is same as the $\pi$ wave function.
The parameter $\zeta$ is either $+1$ or $-1$ depending on the
assignment of the momentum fraction $x$. The $\eta_s=s\bar s$
component of the wave function can be defined in the same way.

The expressions of the relevant distribution amplitudes (DA's) of K meson are the
following \cite{ball98}:
\begin{eqnarray}
\pka(x) &=&  \frac{f_K}{2\sqrt{2N_c} }  6x (1-x)
    \left[1+a_1^{K}C^{3/2}_1(t)+a^{K}_2C^{3/2}_2(t)+a^{K}_4C^{3/2}_4(t)
  \right],\label{piw1}\\
 \pkp(x) &=&   \frac{f_K}{2\sqrt{2N_c} }
   \left\{ 1+(30\eta_3-\frac{5}{2}\rho^2_{K})C^{1/2}_2(t)
   -3\left[ \eta_3\omega_3+\frac{9}{20}\rho^2_K
   (1+6a^K_2)\right]C^{1/2}_4(t)\right\}, \ \ \\
 \pkt(x) &=&  - \frac{f_K}{2\sqrt{2N_c} } t
   \left[ 1+6(5\eta_3-\frac{1}{2}\eta_3\omega_3-\frac{7}{20}\rho^2_{K}
   -\frac{3}{5}\rho^2_Ka_2^{K})
   (1-10x+10x^2)\right] ,\quad\quad\label{piw}
\end{eqnarray}
with the mass ratio $\rho_K=m_K/m_{0K}$.  The Gegenbauer moments can be
given as \cite{ball98}:
\beq
a^K_1=0.2 ,\quad a^K_2=0.25, \quad a^K_4=-0.015.
\eeq
The values of other parameters are $\eta_3=0.015$ and $\omega=-3.0$.
At last the Gegenbauer polynomials $C^{\nu}_n(t)$ are given as:
\beq
C^{1/2}_2(t)&=&\frac{1}{2}(3t^2-1), \qquad C^{1/2}_4(t)=\frac{1}{8}(3-30t^2+35t^4), \non
C^{3/2}_1(t)&=&3t, \qquad C^{3/2}_2(t)=\frac{3}{2}(5t^2-1),\non
C^{3/2}_4(t)&=&\frac{15}{8}(1-14t^2+21t^4),
\label{eq:c124}
\eeq
with $t=2x-1$.

In the quark-flavor mixing scheme, the physical states $\eta$ and $\etar$ are
related to the flavor states $\eta_q= (u\bar u +d\bar d)/\sqrt{2}$  and $\eta_s=s\bar{s}$
through a single mixing angle $\phi$,
\beq
\left(\begin{array}{c} \eta \\ \eta^{\prime} \end{array} \right)
=\left(\begin{array}{cc}
 \cos{\phi} & -\sin{\phi} \\
 \sin{\phi} & \cos{\phi} \\ \end{array} \right)
 \left(\begin{array}{c} \eta_q \\ \eta_s \end{array} \right)
=\left(\begin{array}{c}
 F_1(\phi) (u\bar{u} + d\bar{d})  + F_2(\phi) \; s\bar{s} \\
 F_1^\prime(\phi) (u\bar{u} + d\bar{d})  + F_2^\prime(\phi) \; s\bar{s} \\
\end{array} \right)
\label{eq:e-ep}
\eeq
with
\beq
F_1(\phi)&=& \frac{\cos\phi}{\sqrt{2}}, \quad F_2(\phi)= - \sin\phi, \non
F_1^\prime(\phi)&=& \frac{\sin\phi}{\sqrt{2}}, \quad F_2^\prime(\phi)= \cos\phi.
\label{eq:f1f2phi}
\eeq

The relation between the decay constants $(f_\eta^q, f_\eta^s,f_{\etar}^q,f_{\etar}^s)$ and
$(f_q,f_s,)$ can be written as
\beq
\left(\begin{array}{cc} f^q_\eta & f^s_\eta\\
                        f^q_{\etar} & f^s_{\etar} \end{array} \right)
=\left(\begin{array}{cc}
 \cos{\phi} & -\sin{\phi} \\
 \sin{\phi} & \cos{\phi} \\ \end{array} \right)
 \left(\begin{array}{cc} f_q & 0 \\ 0& f_s \end{array} \right),
 \label{eq:op}
\eeq

The chiral enhancement $m_0^q$ and $m_0^s$ associated
with the two-parton twist-3 $\eta_q$ and $\eta_s$ meson
distribution amplitudes have been defined as \cite{nlo05}
 \beq
 m^{q}_{0}&=&\frac{m^2_{qq}}{2m_q}=\frac{1}{2m_q}[m^2_{\eta}\cos^2\phi
 +m^2_{\eta^{\prime}}\sin^2\phi-\frac{\sqrt{2}f_s}{f_q}(m^2_{\eta^{\prime}}-m^2_{\eta})
 \cos\phi\sin\phi],\label{eq:m0q}\\
m^{s}_{0}&=&\frac{m^2_{ss}}{2m_s}=\frac{1}{2m_s}[m^2_{\eta^{\prime}}\cos^2\phi
 +m^2_{\eta}\sin^2\phi-\frac{\sqrt{2}f_q}{f_s}(m^2_{\eta^{\prime}}-m^2_{\eta})
 \cos\phi\sin\phi],\label{eq:m0s}
 \eeq
by assuming the exact isospin symmetry $m_q=m_u=m_d$.
The three input parameters $f_q, f_s$ and $\phi$ have been
extracted from the data of the relevant exclusive processes \cite{fks98}:
\beq
f_q=(1.07\pm 0.02)f_{\pi},\quad f_s=(1.34\pm 0.06)f_{\pi},\quad \phi=39.3^\circ\pm 1.0^\circ,
\eeq

The distribution amplitudes $\phi_{\eta_q}^{A,P,T}$ represent the axial vector, pseudoscalar
and tensor component of the wave function respectively \cite{ball98}.
They are given as:
\begin{eqnarray}
 \phi_{\eta_q}^A(x) &=&  \frac{f_q}{2\sqrt{2N_c} }
    6x (1-x)
    \left[1+a^{\eta_q}_1C^{3/2}_1(2x-1)+a^{\eta_q}_2 C^{3/2}_2(2x-1)
    \right.\non && \left.+a^{\eta_q}_4C^{3/2}_4(2x-1)
  \right],\label{piw11}\\
 \phi_{\eta_q }^P(x) &=&   \frac{f_q}{2\sqrt{2N_c} }
   \left[ 1+(30\eta_3-\frac{5}{2}\rho^2_{\eta_q } )C^{1/2}_2(2x-1)
\right.\non && \left.
   -3\left\{\eta_3\omega_3+\frac{9}{20}\rho^2_{\eta_q }
   (1+6a^{\eta_q }_2)\right\} C^{1/2}_4(2x-1)\right]  ,\\
 \phi_{\eta_q}^T(x) &=&  \frac{f_q}{2\sqrt{2N_c} } (1-2x)
   \left[ 1+6\left (5\eta_3-\frac{1}{2}\eta_3\omega_3
   -\frac{7}{20}\rho^2_{\eta_q}-\frac{3}{5}\rho^2_{\eta_q }a_2^{\eta_q}\right )
   \right. \non && \left.
   \cdot \left (1-10x+10x^2 \right )\right] ,\quad\quad\label{piw4}
 \end{eqnarray}
where $\rho_{\eta_q}=2m_q/m_{qq}$,$a^{\eta_q}_1=a_1^\pi=0$, $a^{\eta_q}_2=a_2^\pi=0.44\pm 0.22$,
$a^{\eta_q}_4=a_4^\pi=0.25$, and the Gegenbauer polynomials $C^{\nu}_n(t)$ have been
given in Eq.~(\ref{eq:c124}). 
As to the wave function and the corresponding DA's of the $s\bar{s}$ components, we
also use the same form as $q\bar{q}$ but with some parameters changed: $\rho_{\eta_s}
=2m_s/m_{ss}$, $a^{\eta_s}_i=a^{\eta_q}_i$ for $i=1,2,4$.

The transverse momentum $k_T$ is usually converted to
the $b$ parameter by Fourier transformation. The initial conditions of
leading twist $\phi_i(x)$, $i=B,K,\eta, \etar$, are of non-perturbative
origin, satisfying the normalization
\beq
\int_0^1\phi_i(x,b=0)dx=\frac{1}{2\sqrt{6}}{f_i}\;, \label{no}
\eeq
with $f_i$ the meson decay constant.

\section{Decay amplitudes at leading order}\label{sec:lo-1}

\begin{figure}[tb]
\vspace{-3cm}
\centerline{\epsfxsize=18cm \epsffile{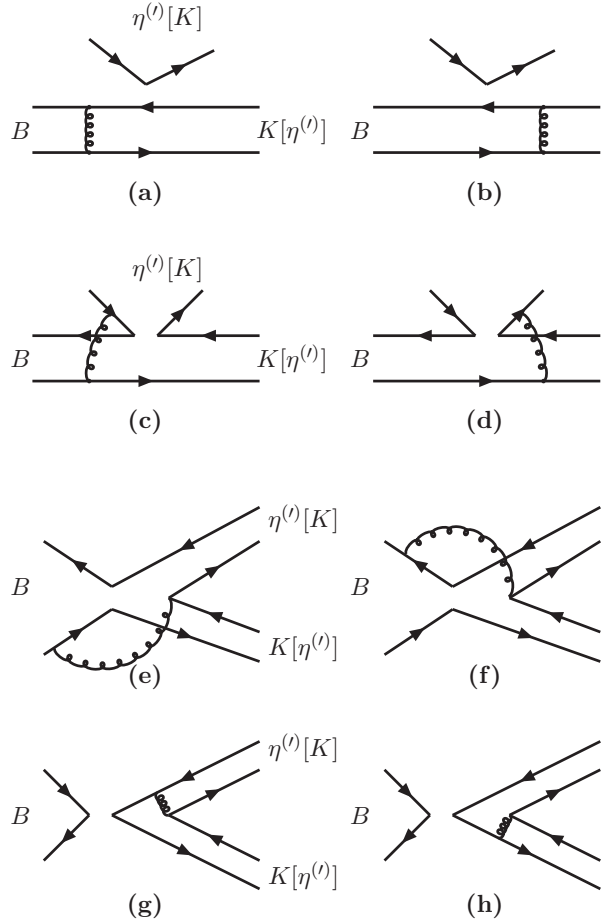}}
\vspace{-10cm}
\caption{ Feynman diagrams which may contribute to the $B\to K \etap$
decays at leading order.}
\label{fig:fig1}
\end{figure}

In  the pQCD approach, the Feynman diagrams  as shown in Fig.~\ref{fig:fig1} may contribute
to $B \to K \etap$ decays at leading order.
As mentioned previously, $B^0 \to K^0 \etap$ decays have been studied in Ref.~\cite{kou02}
by employing the LO pQCD approach. In this section, we firstly calculate the LO decay
amplitudes for four $B \to K \etap$ decays, but in a rather different way to treat the
Feynman diagrams from that in Ref.~\cite{kou02}.

At the leading order in pQCD approach, there are three type
diagrams contributing to the $B \to K \etap $ decays, the factorizable
emission diagrams, the hard-spectator diagrams and the annihilation diagrams,
as illustrated in Fig.\ref{fig:fig1}. From the factorizable
emission diagrams, the corresponding form factors can be extracted by perturbative
calculation. First, we consider the $B\to K \eta$ decay modes, and then extend
the calculation to $B \to K \etar $ decays.

For the usual factorizable emission diagrams 1(a) and 1(b) with the $B \to K$ transition,
i.e., it is the K meson pick up the spectator quark,
the operators $O_1$, $O_2$, $O_{3,4}$ and $O_{9,10}$ are
$(V-A)(V-A)$ currents, the sum of the individual amplitudes is
given as
\beq
F_{eK}&=& \frac{8}{\sqrt{2}}\pi G_F C_F m_B^4\; \int_0^1 d x_{1} dx_{2}\,
\int_{0}^{\infty} b_1 db_1 b_2 db_2\, \phi_B(x_1,b_1) \non & &
\times \left\{ \left[(1+x_2) \pka(\xeba) +(1-2x_2) \rk (\pkp(\xeba)
-\pkt(\xeba))\right]\cdot E_e(t_a) h_e(x_1,x_2,b_1,b_2)\right.\non &&
\left. +2\rk \pkp (\xeba)\cdot E_e(t_a^{\prime})h_e(x_2,x_1,b_2,b_1)
\right\},
\label{eq:ab1}
\eeq
where $\rk=m_0^K/m_B$ with $m_0^K$ is the chiral scale;
$C_F=4/3$ is a color factor, and $\xeba=1-x_2$.
The evolution function $E_e(t)$ and hard function $h_e$ are displayed in
Appendix \ref{sec:aa}. In the above equation, we do not include the
Wilson coefficients of the corresponding operators, which are
process dependent. They will be shown later in the expressions of
total decay amplitude.

Also for diagrams 1(a) and 1(b), the operators $O_{5,6}$ and $O_{7,8}$
have a structure of $(V-A)(V+A)$ currents.
In some decay channels, some of these operators
contribute to the decay amplitude in a factorizable way. Since only
the axial-vector part of $(V+A)$ current contribute to the
pseudo-scaler meson production, $ \langle K |V-A|B\rangle \langle
\etap |V+A | 0 \rangle = -\langle K |V-A |B  \rangle \langle \etap
|V-A|0 \rangle,$ that is
\beq
 F_{eK}^{P1}=-F_{eK}\; .
\label{eq:fekp1}
\eeq
In some other cases, we need to do Fierz transformation  for those
operators to get right color structure for factorization to work.
In this case, we get $(S-P)(S+P)$ operators from $(V-A)(V+A)$ ones.
For these $(S-P)(S+P)$ operators, the corresponding decay amplitude is
\beq
F_{eK}^{P2}&=& \frac{16}{\sqrt{2}}\pi G_FC_F m_B^4 \int_{0}^{1}d x_{1}d
x_{2}\,\int_{0}^{\infty} b_1d b_1 b_2d b_2\, \pb(x_1) \non & &
\times
 \left\{\re \left[ \pka(\xeba)+ \rk((2+x_2) \pkp (\xeba)+x_2\pkt(\xeba))\right]
\cdot E_e(t_a)h_e (x_1,x_2,b_1,b_2)\right. \non & &\left. \
   +2\rk\re\pkp (\xeba)
\cdot E_e(t_a^{\prime})h_e(x_2,x_1,b_2,b_1) \right\} \; .
\label{eq:fek-p2}
\eeq
where $\re=m_0^q /m_B$, and $m_0^q=m_0^{\eta_q}$ is the chiral scale defined
in Eq.~(\ref{eq:m0q}).

For the non-factorizable diagrams 1(c) and 1(d), all three meson
wave functions are involved. The integration of $b_2$ can be
performed using $\delta$ function $\delta(b_3-b_2)$, leaving only
integration of $b_1$ and $b_3$. For the $(V-A)(V-A)$ operators, the
result is
\beq
M_{eK}&=& \frac{16}{\sqrt{3}}\pi G_F C_F m_B^4\;
\int_{0}^{1}d x_{1}d x_{2}\,d x_{3}\,\int_{0}^{\infty} b_1d b_1 b_3d
b_3\, \pb(x_1,b_1) \pea(\xsba) \non
 & &\times \left\{\left[
-\rk x_2\left(\pkp(\xeba)+\pkt(\xeba)\right)+(1-x_3)\pka(\xeba)\right
] \right. \non
 & &\left. \ \ \ \cdot E_e^{\prime}(t_b) h_n(x_1,x_2,1-x_3,b_1,b_3)
\right.\non
&& \left.
+\left[-(x_2+x_3)\pka(\xeba)+\rk
x_2\left(\pkp(\xeba)-\pkt(\xeba)\right)\right]
\right.\non &&
\left. \ \ \
\cdot E_e^{\prime}(t_b^{\prime}) h_n(x_1,x_2,x_3,b_1,b_3) \right\},
\eeq
where $\phi_{\eta}$ denotes $\phi_{\eta_q}$ or $\phi_{\eta_s}$.

There are two kinds of contributions from $(V-A)(V+A)$ operators:
$M_{eK}^{P1}$ and $M_{eK}^{P2}$, corresponding to the $(V-A)(V+A)$ and
$(S-P)(S+P)$ type operators respectively:
\beq
M_{eK}^{P1}&=& \frac{16}{\sqrt{3}} \pi G_F C_F m_B^4\;
\int_{0}^{1}d x_{1}d x_{2}\,d x_{3}\,\int_{0}^{\infty} b_1d b_1
b_3d b_3\, \pb(x_1,b_1)\non
&& \cdot \left\{ \left[ (1-x_3)\pka(\xeba) \left(\pep(\xsba)-\pet(\xsba)\right)
\right. \right.\non
&& \left.\left.
+ \rk(1-x_3)\left(\pkp(\xeba)+\pkt(\xeba)\right)
\left(\pep(\xsba)-\pet(\xsba)\right) \right.\right.\non
&& \left. \left.
+ \rk x_2\left(\pkp(\xeba)-\pkt(\xeba)\right)
 \left(\pep(\xsba)+\pet(\xsba)\right)\right]\right.\non
 && \left.
\ \  \cdot E_e^{\prime}(t_b)h_n(x_1,x_2,1-x_3,b_1,b_3)
 \right. \non
&&\left.
-\left[ \;  x_3\pka(\xeba)\left(\pep(\xsba)+\pet(\xsba)\right)
 \right.\right.\non
&& \left. \left.
+ r_2 x_3 \left(\pkp(\xeba)+\pkt(\xeba)\right)\left(\pep(\xsba)+\pet(\xsba)\right)
 \right.\right.\non
&& \left. \left.
+ r_2 x_2 \left(\pkp(\xeba)-\pkt(\xeba)\right)\left(\pep(\xsba)-\pet(\xsba)\right)\; \right]
\right.\non
&& \left.
\ \ \cdot  E_e^{\prime}(t_b^{\prime}) h_n(x_1,x_2,x_3,b_1,b_3) \right\},
\eeq

\beq
M_{eK}^{P2}&=& \frac{16}{\sqrt{3}}\pi G_F C_F m_B^4 \int_{0}^{1}d
x_{1}d x_{2}\,d x_{3}\,\int_{0}^{\infty} b_1d b_1 b_3d
b_3\,\pb(x_1,b_1)\pea(\xsba)\non &&
\cdot \left\{\left[-(1+x_2-x_3)\pka(\xeba)+x_2\rk\left(\pkp(\xeba)-\pkt(\xeba)\right)\right]
\right.\non
&& \left. \ \ \cdot E_e^{\prime}(t_b)h_n(x_1,x_2,1-x_3,b_1,b_3)
\right.\non && \left.+
\left[x_3\pka(\xeba)-x_2\rk\left(\pkp(\xeba)+\pkt(\xeba)\right)\right]
\cdot E_e^{\prime}(t_b^{\prime})
h_n(x_1,x_2,x_3,b_1,b_3)\right\}.
\eeq

For the non-factorizable annihilation diagrams 1(e) and 1(f), again all three
wave functions
are involved. Here we have two kinds of contributions: $M_{aK}^{P2}=0$,
$M_{aK}$ and $M_{aK}^{P1}$ describe the contributions from the $(V-A)(V-A)$ and
$(V-A)(V+A)$ type operators, respectively,
 \beq
 M_{aK}&=& \frac{16}{\sqrt{3}}\pi G_F C_F m_B^4\; \int_{0}^{1}d x_{1}d x_{2}\,d
x_{3}\,\int_{0}^{\infty} b_1d b_1 b_3d b_3\, \pb(x_1,b_1)\non &&
\cdot \left\{\left[(1- x_2)\pka(\xeba) \pea(\xsba) +\rk \re (1-x_2)
\left(\pkp(\xeba)+\pkt(\xeba) \right) \left(\pep(\xsba)-\pet(\xsba)
\right)\right.\right.\non && \left.\left.
 + \rk\re x_3\left(\pkp(\xeba)-\pkt(\xeba)
\right)\left(\pep(\xsba)+\pet(\xsba)\right)
 \right]\cdot E_a^{\prime}(t_c)h_{na}(x_1,x_2,x_3,b_1,b_3)\right.\non &&
\left.   -\left[ x_3 \pka(\xeba) \pea(\xsba) +4\rk \re
\pkp(\xeba)\pep(\xsba)-
 \rk\re(1-x_3)\left(\pkp(\xeba)+\pkt(\xeba)\right)\right.\right.\non &&
 \cdot\left.\left.\left(\pep(\xsba)-
 \pet(\xsba)\right)- \rk\re x_2
 \left(\pkp(\xeba)-\pkt(\xeba)\right)\left(\pep(\xsba)+\pet(\xsba)\right)\right]
\right.\non &&\times\left.
\ \ \cdot E_a^{\prime}(t_c^{\prime})h_{na}(x_1,x_2,x_3,b_1,b_3)
 \right \}\; ,
 \eeq
\beq
M_{aK}^{P1}&=& \frac{16}{\sqrt{3}}\pi G_F C_F m_B^4\int_{0}^{1}d
x_{1}d x_{2}\,d x_{3}\,\int_{0}^{\infty} b_1d b_1 b_3d b_3\,
\pb(x_1,b_1)
 \non
& &\times \left\{ \left[ -(1-x_2) \rk
\pea(\xsba)\left(\pkp(\xeba)+\pkt(\xeba) \right)+\re x_3
\pka(\xeba)\left( \pep(\xsba) -\pet(\xsba)\right)\right]\right.\non
&&\left.\times
E_a^{\prime}(t_c)h_{na}(x_1,x_2,x_3,b_1,b_3)-\left[(x_2+1)\rk
\pea(\xsba)\left(\pkp(\xeba)+ \pkt(\xeba) \right) \right.\right.\non
&&\left.\left.+\re
(x_3-2)\pka(\xeba)\left(\pep(\xsba)-\pet(\xsba)\right)\right]
E_a^{\prime}(t_c^{\prime})h_{na}(x_1,x_2,x_3,b_1,b_3) \right \} \; .
\eeq

The factorizable annihilation diagrams 1(g) and 1(h) involve only $K$
and $\etap$ wave functions. There are also three kinds of decay
amplitudes for these two diagrams. $F_{aK}$, $F_{aK}^{P1}$ and
$F_{aK}^{P2}$:
\beq
F_{aK}&=&F_{aK}^{P1}= \frac{8}{\sqrt{2}} \pi G_F C_F m_B^4\;  \int_{0}^{1}dx_{2}\,d x_{3}\,
\int_{0}^{\infty} b_2d b_2b_3d b_3 \, \left\{-\left[ (1-x_2)
\pka(\xeba) \pea(\xsba)\right.\right.\non &&\left.\left. +4 \re
\rk\pkp(\xeba)\pep(\xsba)-2\rk\re
x_2\pep(\xsba)\left(\pkp(\xeba)+\pkt(\xeba)\right)\right]
\right.\non && \left.
\ \ \cdot E_a(t_d)h_a(x_3,1-x_2,b_3,b_2)
\right.\non
&&\left.
+ \left[ x_3 \pka(\xeba) \pea(\xsba)
+2 \re \rk \pkp(\xeba) \left(\pep(\xsba)+\pet(\xsba)\right)
\right. \right.\non
&&\left.\left.
+ 2\re\rk x_3\pkp(\xeba)\left(\pep(\xsba)-\pet(\xsba)\right)\right]\cdot
 E_a(t_d^{\prime}) h_a(1-x_2,x_3,b_2,b_3) \right \}\; ,
\eeq
\beq
F_{aK}^{P2}&=& \frac{16}{\sqrt{2}} \pi G_F C_F m_B^4 \int_{0}^{1}d
x_{2}\,d x_{3}\,\int_{0}^{\infty} b_2d b_2b_3d b_3 \,\non &&
\cdot \left\{ \left[ \rk (1-x_2)
\left(\pkp(\xeba)-\pkt(\xeba)\right)\pea(\xsba)+2\re \pka(\xeba)
\pep(\xsba) \right]\right.
 \non
&&\left. \ \ \cdot E_a(t_d) h_a(x_3,1-x_2,b_3,b_2)\right.
 \non
&&\left.+\left[2\rk
\pkp(\xeba)\pea(\xsba)+x_3\re\pka(\xeba)(\pep(\xsba)+\pet(\xsba))\right]
\right.\non &&\left.
\ \ \cdot E_a(t_d^{\prime})
h_a(1-x_2,x_3,b_2,b_3)\right\}\;
\label{eq:gh}.
\eeq
The evolution function $E_i(t_j)$ and hard function $h_i$ appeared in
Eqs.~(\ref{eq:fek-p2}-\ref{eq:gh}) are given explicitly in Appendix \ref{sec:aa}.

If we exchange the $K$ and $\etap$ in Fig.~1, the corresponding
decay amplitudes for new diagrams will be similar with
those as given in Eqs.(\ref{eq:ab1}-\ref{eq:gh}), since the $K$
and $\etap$ are all pseudoscalar mesons and have the similar wave
functions. The decay amplitudes for new diagrams,
say $F_{e\eta}$, $F_{e\eta}^{P1,P2}$, $M_{e\eta}$, $M_{e\eta}^{P1,P2}$,
$M_{a\eta}$,$M_{a\eta}^{P1}$, $F_{a\eta}$,$F_{a\eta}^{P1,P2}$,
can be obtained from those those as given in Eqs.(\ref{eq:ab1}-\ref{eq:gh})
by the following replacements
\beq
\pka \leftrightarrow
\phi^{A}_{\eta^{(\prime)}}, \quad \pkp \leftrightarrow
\phi^{P}_{\eta^{(\prime)}} , \quad \pkt \leftrightarrow
\phi^{T}_{\eta^{(\prime)}}, \quad \rk \leftrightarrow
r_{\eta^{(\prime)}}.
\label{repalce1}
\eeq

For $B^0 \to K^0 \eta$ decay, by combining the contributions from all possible
configuration of Feynman diagrams, one finds the total decay
amplitude with the inclusion of the corresponding Wilson coefficients
as follows
\beq
{\cal M}(K^0 \eta) &=&< K^0 \eta| H_{eff}|B^0>
= F_{eK} \left \{\left[ \xi_u a_2-\xi_t
\left(2a_3-2a_5-\frac{1}{2}a_7+\frac{1}{2}a_9\right)\right ]
f_{\eta}^q  \right.\non &&\left.
-\xi_t\left(a_3+a_4-a_5+\frac{1}{2}a_7-\frac{1}{2}a_9-\frac{1}{2}a_{10}\right)
f^s_{\eta}  \right \}\non &&
-F_{e\eta}\xi_t\left(a_4-\frac{1}{2}a_{10}\right)f_K F_1(\phi)\non
&& -\left[F_{eK}^{P_2}f^s_{\eta} +F_{e\eta}^{P_2}f_K
F_1(\phi)\right] \xi_t \left (a_6 -\frac{1}{2}a_8\right)\non
&&
-\left[ F_{ak}F_2(\phi)+F_{a\eta}F_1(\phi)\right]\; \xi_t \;
\left(a_4-\frac{1}{2}a_{10} \right ) \non
&&
+\left[ F_{aK}^{P_2}F_2(\phi)+F_{a\eta}^{P_2}F_1(\phi)\right] \; \xi_t \;
\left(a_6-\frac{1}{2}a_8\right)\; f_B \non
&&
+ M_{eK} \left \{ \left[ \xi_u C_2 -\xi_t \cdot
\left(2C_4+\frac{1}{2}C_{10}\right)\right ]F_1(\phi)\right.\non
&& \left.
-\xi_t \left ( C_3+C_4 -\frac{1}{2}C_{9}-\frac{1}{2}C_{10}\right )F_2(\phi)
\right\}\non &&
-M_{e\eta}\xi_t\left(C_3-\frac{1}{2}C_9\right)F_1(\phi)
-\left[M_{eK}^{P1}F_2(\phi)+M_{e\eta}^{P1}F_1(\phi)\right]\xi_t\left(C_5-\frac{1}{2}C_7\right)\non
&&
-M_{eK}^{P_2}\xi_t\left[\left(2C_6+\frac{1}{2}C_8\right)F_1(\phi)+(C_6-\frac{1}{2}C_8)F_2(\phi)\right]
\label{eq:k0eta}
\eeq
where $\xi_u = V_{ub}^*V_{us}$, $\xi_t = V_{tb}^*V_{ts}$,
and $F_1(\phi), F_2(\phi)$ are the mixing factors as given in Eq.~(\ref{eq:f1f2phi}).

The coefficients $a_i$ in Eq.~(\ref{eq:k0eta}) are the combinations of the Wilson coefficients
$C_i$, and  have been defined as usual
\beq
a_1&=& C_2 + \frac{C_1}{3}, \quad a_2= C_1 + \frac{C_2}{3}, \non
a_{i}&=& C_i + \frac{C_{i+1}}{3}, \ \ {\rm for} \ \ i=3,5,7,9,\non
a_{i}&=& C_i + \frac{C_{i-1}}{3}, \ \ {\rm for} \ \ i=4,6,8,10.
\label{eq:aimu}
\eeq

Similarly, the decay amplitude for $B^+ \to K^+ \eta$ can be written as
\beq
{\cal M}(K^+ \eta) &=& < K^+ \eta| H_{eff}|B^0>\non
&=&
F_{eK} \left \{\left[\xi_u a_2-\xi_t
\left(2a_3-2a_5-\frac{1}{2}a_7+\frac{1}{2}a_9\right)\right ]
f_\eta^q  \right.\non &&\left.
-\xi_t\left(a_3+a_4-a_5+\frac{1}{2}a_7-\frac{1}{2}a_9-\frac{1}{2}a_{10}\right)
f^s_{\eta}  \right\}\non &&
+\left[F_{e\eta}F_1(\phi)f_K+\left(F_{a\eta}F_1(\phi)+F_{aK}F_2(\phi)\right)f_B\right]
\xi_u a_1\non &&
-\left[F_{e\eta}F_1(\phi)f_K+\left(F_{a\eta}F_1(\phi)+F_{aK}F_2(\phi)\right)f_B\right]
\xi_t\left(a_4+a_{10}\right)\non &&
-\left[F^{P2}_{e\eta}F_1(\phi)f_K+\left(F^{P2}_{a\eta}F_1(\phi)
+F_{aK}^{P2}F_2(\phi)\right)f_B\right]\xi_t\left(a_6+a_8\right)\non
&& -F^{P2}_{eK}f_\eta^s\xi_t
\left(a_6-\frac{1}{2}a_8\right)-M_{eK}^{P1}\xi_t\left(C_5-\frac{1}{2}C_7\right)\non
&& +M_{eK}\left\{\left[\xi_u
C_2-\xi_t\left(2C_4+\frac{1}{2}C_{10}\right)\right]
F_1(\phi)\right.\non &&\left. -\xi_t\left(C_3+C_4-\frac{1}{2}
C_9-\frac{1}{2}C_{10}\right)F_2(\phi)\right\}\non &&
+\left[M_{aK}F_2(\phi)+\left(M_{e\eta}+M_{a\eta}\right)F_1(\phi)\right]
\left[\xi_u C_1-\xi_t\left(C_3+C_9\right)\right]\non &&
-\left[M^{P1}_{aK}F_2(\phi)+\left(M_{e\eta}^{P1}+M_{a\eta}^{P1}
\right)F_1(\phi)\right]\xi_t(C_5+C_7) \non &&
-M_{eK}^{P2}\xi_t\left[\left(2C_6+\frac{1}{2}C_8\right)F_1(\phi)
+\left(C_6-\frac{1}{2}C_8 \right)F_2(\phi)\right].
\label{eq:kpeta}
\eeq

The total decay amplitude for $B^0 \to K^0 \etar$ and $B^+ \to K^+ \etar $ can be
obtained easily from Eqs.(\ref{eq:k0eta}) and
(\ref{eq:kpeta}) by the following replacements
\beq
f_\eta^{d} &\to& f_{\eta^\prime}^d, \quad f_\eta^s \to f_{\eta^\prime}^s,\non
F_1(\phi) &\to& F'_1(\phi), \quad  F_2(\phi) \to F'_2(\phi).
\label{eq:f1f2phip}
\eeq

\section{NLO contributions in pQCD approach}\label{sec:nlo}

\subsection{General discussion}\label{sec:4-0}

\begin{figure}[tb]
\vspace{-5cm} \centerline{\epsfxsize=18 cm \epsffile{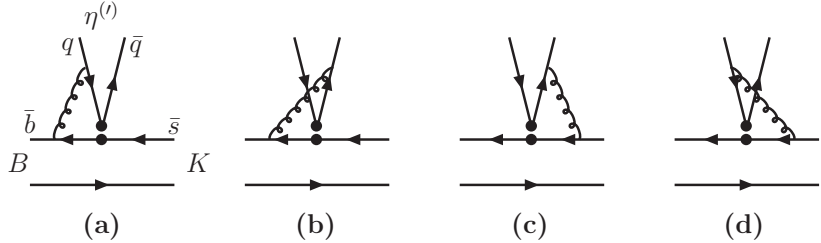}}
\vspace{-18cm}
\caption{NLO vertex corrections to the factorizable amplitudes.}
\label{fig:fig2}
\end{figure}

\begin{figure}
\vspace{-4cm}
\centerline{\epsfxsize=18 cm \epsffile{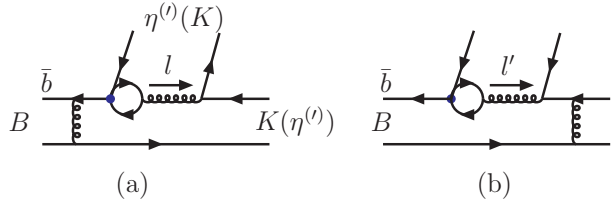}}
\vspace{-18cm}
\caption{Quark-loop amplitudes.}
\label{fig:fig3}
\end{figure}

\begin{figure}[tb]
\vspace{-5cm}
\centerline{\epsfxsize=18 cm \epsffile{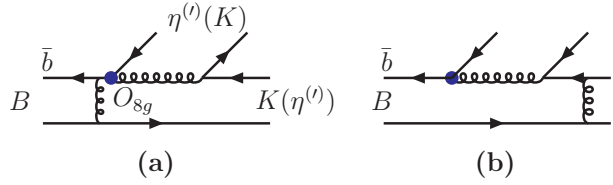}}
\vspace{-18cm}
\caption{Chromo-magnetic penguin amplitudes ($O_{8g}$). There are nine relevant
Feynman diagrams as shown in Ref.~\cite{o8g2003}.
Here we show the first two only, which
provide dominant contribution of such diagrams. }
\label{fig:fig4}
\end{figure}

\begin{figure}[tb]
\vspace{-5cm}
\centerline{\epsfxsize=18 cm \epsffile{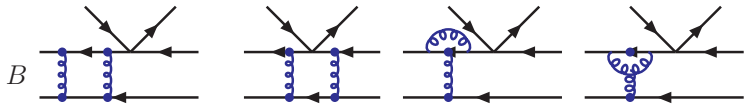}}
\vspace{-18cm}
\caption{The four typical Feynman diagrams, which contributes to the form factors at
NLO level.}
\label{fig:fig5}
\end{figure}

\begin{figure}[thb]
\vspace{-5cm}
\centerline{\epsfxsize=18 cm \epsffile{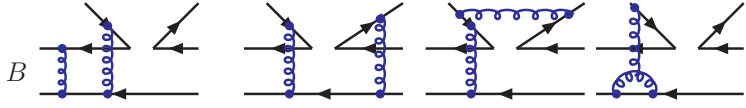}}
\vspace{-18cm}
\caption{The four typical hard-spectator Feynman diagrams, which contributes
at NLO level.}
\label{fig:fig6}
\end{figure}

\begin{figure}[thb]
\vspace{-5cm}
\centerline{\epsfxsize=18 cm \epsffile{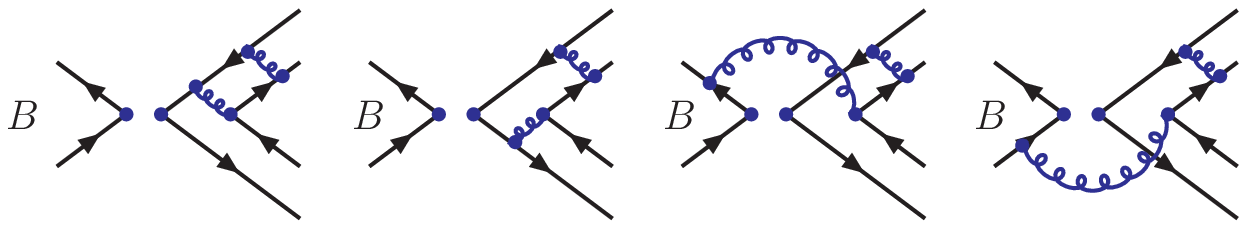}}
\vspace{-18cm}
\caption{The four typical annihilation Feynman diagrams, which contributes
at NLO level.}
\label{fig:fig7}
\end{figure}

The power counting in the pQCD factorization approach \cite{nlo05} is different
from that in the QCD factorization\cite{bbns99,bn651}.
When compared with the previous LO calculations in pQCD \cite{li2003,xiao06,xiao07},
the following NLO contributions should be considered:
\begin{enumerate}
\item
The LO Wilson coefficients $C_i(\mw)$ will be replaced by those at NLO level in NDR scheme
\cite{buras96}, and the NLO  RG evolution
matrix $U(t,m,\alpha)$ instead of $U(m_1,m_2)^{(0)}$,
as defined in Ref.~\cite{buras96}, will be used here:
\beq
U(m_1,m_2,\alpha) = U(m_1,m_2) + \frac{\alpha}{4\pi} R(m_1,m_2)
\label{eq:um1m2a}
\eeq
where the function  $U(m_1,m_2)$ and $R(m_1,m_2)$ represent the QCD and QED evolution and
have been defined in Eq.~(6.24) and (7.22) in Ref.~\cite{buras96}.
We also introduce a cut-off $\mu_0=1.0$ GeV for the QCD running of $C_i(t)$
in the final integration.

\item
The strong coupling constant $\alpha_s(t)$ at two-loop level as given in Eq.~(\ref{eq:asnlo})
will be used.

\item
Besides the LO hard kernel $H^{(0)}(\alpha_s)$, the NLO hard kernel
$H^{(1)}(\alpha_s^2)$ should be included.
All the Feynman diagrams,  which lead to the decay amplitudes proportional to
$\alpha^2_s(t)$, should be considered.
Such Feynman diagrams can be grouped into following classes:

\begin{itemize}
\item[]{I:}
The vertex corrections, as illustrated in Fig.~\ref{fig:fig2},
the same set as that studied in the QCDF approach.

\item[]{II:}
The NLO contributions from quark-loops, as illustrated  in Fig.~\ref{fig:fig3}.

\item[]{III:}
The NLO contributions from chromo-magnetic penguins, i.e. the operator $O_{8g}$,
as illustrated  in Fig.~\ref{fig:fig4}. There are totally nine relevant Feynman diagrams
as given in Ref.~\cite{o8g2003}, if the Feynman diagrams involving
three-gluon vertex are also included.
We here show the first two only, and they provide
the dominant NLO contributions, according to Ref.~\cite{o8g2003}.

\item[]{IV:}
The NLO contributions to the Feynman diagrams (1a,1b) corresponding to the
extraction of from factors, as illustrated  in Fig.~\ref{fig:fig5}.
There are totally 13 relevant Feynman diagrams, we here show four of them only.

\item[]{V: }
The NLO contributions to the hard-spectator Feynman diagrams (1c,1d), as illustrated
in Fig.~\ref{fig:fig6}. There are totally 56 relevant Feynman diagrams,
we here show four only.

\item[]{VI: }
The NLO contributions to the annihilation Feynman diagrams (1e,1h), as illustrated
in Fig.~\ref{fig:fig7}. we here show only four such diagrams.

\end{itemize}
\end{enumerate}

For the last four classes (III-VI), the Feynman diagrams involving three-gluon vertex
should be included.
At present, the calculations for the vertex corrections, the quark-loops and
chromo-magnetic penguins have been available and will be considered here.
For the Feynman diagrams as shown in Figs.~5-7, however,
the analytical calculations have not been completed yet.
What we can do here is to include the NLO contributions to the hard kernel $H$.

\subsection{Vertex corrections}\label{sec:vc2}

The vertex corrections to the factorizable emission diagrams, as illustrated by
Fig.~\ref{fig:fig2}, have been calculated years ago in the QCD factorization
appeoach\cite{bbns99,bn651,npb675}.

For the emission diagram, there are 4 kinds of single
gluon exchange responsible for the effective vertex as labeled in
Fig.\ref{fig:fig2}. The contributions from the soft gluons and
collinear gluons are power suppressed, that is to say the total
contributions of these four figures are infrared finite. For
charmless B meson decays, these corrections can be calculated without
considering the transverse momentum effects of the quark at the
end-point in collinear factorization theorem.
Therefore, there is no need to employ the $k_T$ factorization theorem.
In fact, the difference of the calculations induced by
considering or not considering the parton transverse momentum is rather small \cite{nlo05},
say less than $10\%$, and therefore can be neglected.
Consequently, one can use the vertex corrections as given in Ref.~\cite{npb675} directly.
The vertex corrections can then be absorbed into the re-definition of the Wilson coefficients
$a_i(\mu)$ by adding a vertex-function $V_i(M)$ to them
\cite{bbns99,npb675}
\beq
a_i(\mu)&\to & a_i(\mu) +\frac{\alpha_s(\mu)}{4\pi}C_F\frac{C_i(\mu)}{3} V_i(M),\ \ for
\ \ i=1,2; \non
a_j(\mu)&\to & a_j(\mu)+\frac{\alpha_s(\mu)}{4\pi}C_F\frac{C_{j+1}(\mu)}{3}
V_j(M), \ \  for \ \ j=3,5,7,9, \non
a_j(\mu)&\to & a_j(\mu)+\frac{\alpha_s(\mu)}{4\pi}C_F\frac{C_{j-1}(\mu)}{3}
V_j(M), \ \  for \ \ j=4,6,8,10,
\label{eq:aimu-2}
\eeq
where M is the meson emitted from the weak vertex. When $M$ is a
pseudo-scalar meson, the vertex functions $V_{i}(M)$ are given ( in the NDR
scheme) in Refs.~\cite{nlo05,npb675}:
\beq
V_i(M)&=&\left\{ \begin{array}{cc}
12\ln\frac{m_b}{\mu}-18+\frac{2\sqrt{6}}{f_M}\int_{0}^{1}dx\phi_M^A(x)g(x),
& {\rm for}\quad i= 1-4,9,10,\\
-12\ln\frac{m_b}{\mu}+6-\frac{2\sqrt{6}}{f_M}\int_{0}^{1}dx\phi_M^A(x)g(1-x),
&{\rm for}\quad i= 5,7,\\
-6+\frac{2\sqrt{6}}{f_M}\int_{0}^{1}dx\phi_M^P(x)h(x),
&{\rm for} \quad i= 6,8,\\
\end{array}\right.
\label{eq:vim}
\eeq
where $f_M$ is the decay constant of the meson M; $\phi_M^A(x)$ and $\phi_M^P(x)$ are
the twist-2 and twist-3 distribution amplitude of the meson M, respectively.
The hard-scattering
functions $g(x)$ and $h(x)$ in Eq.~(\ref{eq:vim}) are:
\beq
g(x)&=& 3\left(\frac{1-2x}{1-x}\ln x-i\pi\right)\non
&& +\left[2Li_2(x)-\ln^2x+\frac{2\ln
x}{1-x}-\left(3+2i\pi\right)\ln x -(x\leftrightarrow 1-x)\right],\\
h(x)&=&2Li_2(x)-\ln^2x-(1+2i\pi)\ln x-(x\leftrightarrow 1-x),
\eeq
where $Li_2(x)$ is the dilogarithm function. As shown in Ref.~\cite{nlo05},
the $\mu$-dependence of the Wilson coefficients $a_i(\mu)$ will be improved generally
by the inclusion of the vertex corrections.


\subsection{Quark loops}\label{sec:ql02}

The contribution from the so-called ``quark-loops" is a kind of penguin correction
with the four quark operators insertion, as illustrated by Fig.~\ref{fig:fig3}.
In fact this is generally called the BSS mechanism\cite{bss79}, which
provide the strong phase needed to induce the CP violation in QCDF approach.
We here include quark-loop amplitude from the
operators $O_{1,2}$ and $O_{3-6}$ only. The quark loops from $O_{7-10}$
will be neglected due to their smallness.

For the $b\to s$ transition, the contributions from the various
quark loops are given by:
\beq
H_{eff}^{(ql)}&=&-\sum\limits_{q=u,c,t}\sum\limits_{q{\prime}}\frac{G_F}{\sqrt{2}}
V_{qb}V_{qs}^{*}\frac{\alpha_s(\mu)}{2\pi}C^{q}(\mu,l^2)\left(\bar{s}\gamma_\rho
\left(1-\gamma_5\right)T^ab\right)\left(\bar{q}^{\prime}\gamma^\rho
T^a q^{\prime}\right),
\eeq
where $l^2$ is  the invariant mass of the
gluon, which attaches the quark loops in Fig.\ref{fig:fig3}. The
functions $C^{q}(\mu,l^2)$ are written as
\beq
C^{q}(\mu,l^2)&=&\left[G^{q}(\mu,l^2)-\frac{2}{3}\right]C_2(\mu),
\label{eq:qlc}
\eeq
for $q=u,c$ and
\beq
C^{(t)}(\mu,l^2)&=&\left[G^{(s)}(\mu,l^2)-\frac{2}{3}\right]
C_3(\mu)+\sum\limits_{q{\prime\prime}=u,d,s,c}G(q^{\prime\prime})(\mu,l^2)
\left[C_4(\mu)+C_6(\mu)\right].\label{eq:eh}
\eeq
The function $G^{(c)}(\mu,l^2)$ for the loop of the massive
$q(q=u,d,s,c)$ quark is given by \cite{nlo05}
\beq
G^{(q)}(\mu,l^2)&=&-4\int_{0}^{1}dx x(1-x)\ln\frac{m_q^2-x(1-x)l^2}{\mu^2}
\eeq
$m_q$ is the possible quark mass. The explicit  expressions of the function
$G^{(q)}(\mu,l^2)$ after the integration can be found, for example, in
Ref.~\cite{nlo05}.

It is straightforward to calculate the decay amplitude for Fig.\ref{fig:fig3}a and
\ref{fig:fig3}b. We find
two kinds of topological decay amplitudes:
\beq
M^{(q)}_{K\eta_s}&=& -\frac{8}{\sqrt{6}} C_F^2 m_B^4\;  \int_{0}^{1}d
x_{1}d x_{2}\,d x_{3}\,\int_{0}^{\infty} b_1d b_1 b_2d
b_2\,\phi_B(x_1,b_1)\non
&& \cdot \left\{\left[\left(1+x_2\right) \pka(\xeba)\pesa(\xsba)
+\rk\left(1-2x_2\right)\left(\pkp(\xeba)-\pkt(\xeba)\right)\pesa(\xsba)
\right.\right.\non &&
\left.\left.
+2\res\pka(\xeba)\pesp(\xsba)
+2\rk\res\left(\left(2+x_2\right)\pkp(\xeba)+
x_2\pkt(\xeba)\right)\pesp(\xsba)\right]
\right.\non &&
\left.
\ \ \cdot E^{(q)}(t_q,l^2) h_e(x_2,x_1,b_2,b_1)
\right.\non &&
\left.
+ \left[ 2\rk\pkp(\xeba)\pesa(\xsba)
+4\rk\res\pkp(\xeba)\pesp(\xsba)\right]
\right.\non &&
\left.
\ \ \cdot E^{(q)}(t^{\prime}_q,l^{\prime 2 }) h_e(x_1,x_2,b_1,b_2)\right\},
\label{eq:mqketa}
\eeq
for $B \to K $ transition, and
\beq
M^{(q)}_{\eta_q K}&=& - \frac{8}{\sqrt{6}} C_F^2 m_B^4\;
\int_{0}^{1}d x_{1}d x_{2}\,d x_{3}\,\int_{0}^{\infty} b_1d b_1 b_2d
b_2\,\phi_B(x_1,b_1)\non
&& \cdot \left\{\left[\left(1+x_2\right) \peqa(\xeba)\pka(\xsba)
+\re \left(1-2x_2\right)\left(\peqp(\xeba)-\peqt(\xeba)\right)\pka(\xsba)
\right.\right.\non &&
\left.\left.
+2\rk \peqa(\xeba)\pkp(\xsba)
+2\re\rk \left(\left(2+x_2\right)\peqp(\xeba)+
x_2 \peqt(\xeba)\right)\pkp(\xsba)\right]
\right.\non &&
\left.
\ \ \cdot E^{(q)}(t_q,l^2) h_e(x_2,x_1,b_2,b_1)
\right.\non &&
\left.
+ \left[ 2\re \peqp(\xeba)\pka(\xsba)
+4\re\rk\peqp(\xeba)\pkp(\xsba)\right]
\right.\non &&
\left.
\ \ \cdot E^{(q)}(t^{\prime}_q,l^{\prime 2}) h_e(x_1,x_2,b_1,b_2)\right\},
\label{eq:mqetak}
\eeq
for $B \to \eta$ transition. Here $\re =m_0^q /m_B$ and $\res=m_0^s/m_B$.
The evolution factors in Eqs.~(\ref{eq:mqketa},\ref{eq:mqetak}) take the form of
\beq
E^{(q)}(t,l^2)&=& C^{(q)}(t,l^2)\;\alpha_s^2(t)\cdot \exp\left [ -S_{ab} \right ],
\label{eq:eqtl2}
\eeq
with the Sudakov factor $S_{ab}$ and the hard function $h_e(x_1,x_2,b_1,b_2) $
as given in Eq.~(\ref{eq:sab}) and Eq.~(\ref{eq:he3}) respectively, and finally
the hard scales and the gluon invariant masses are
\beq
t_{q}&=& {\rm max}(\sqrt{x_2}m_B,\sqrt{x_1 x_2}m_B,\sqrt{(1-x_2) x_3}m_B,1/b_1,1/b_2);,\non
t_{q}^{\prime}&=& {\rm max}(\sqrt{x_1}m_B,\sqrt{x_1 x_2}m_B,\sqrt{|x_3-x_1|}m_B,1/b_1,1/b_2),
\label{eq:tq-tqp}\\
l^2          &=& (1-x_2) x_3 m_B^2 - |\bfkk_{\rm 2T} -\bfkk_{\rm 3T} |^2
\approx (1-x_2) x_3 m_B^2 , \non
l^{\prime 2} &=& (x_3-x_1) m_B^2- |\bfkk_{\rm 1T} -\bfkk_{\rm 3T} |^2 \approx (x_3-x_1) m_B^2.
\label{eq:l2lp2}
\eeq

For $B \to K \etar$ decays, we find  the same decay amplitude.
Finally, the total ``quark-loop" contribution to the considered $B \to K \etap$ ($K=K^0, K^+$)
decays can be written as
\beq
M_{K\eta}^{(ql)} &=&  <K\eta|{\cal H}_{eff}^{ql}|B>
= \frac{G_F}{\sqrt{2}}\; \sum_{q=u,c,t} \lambda_q \; \left [ M^{(q)}_{K\eta_s}\; F_2(\phi)
+ M^{(q)}_{\eta_q K}\; F_1(\phi) \right],
\label{eq:mketa}\\
M_{K\etar}^{(ql)} &=&  <K\etar|{\cal H}_{eff}^{(ql)}|B>
= \frac{G_F}{\sqrt{2}}\;
\sum_{q=u,c,t} \lambda_q \; \left [ M^{(q)}_{K\eta_s}\; F_2^\prime(\phi)
+ M^{(q)}_{\eta_q K}\; F_1^\prime(\phi) \right],
\label{eq:mketap}
\eeq
where $\lambda_q = V_{qb}V_{qs}^{*}$.  The mixing parameters
$F_1(\phi)$, $F_1^\prime(\phi)$,
$F_2(\phi)$ and $F_2^\prime(\phi)$ have been defined in Eqs.~(\ref{eq:f1f2phi}).

It is note that the quark-loop corrections are mode dependent. The
assumption of a constant gluon invariant mass in FA introduces a
large theoretical uncertainty as making predictions. In the pQCD approach,
however, the gluon invariant mass is related to the parton momenta unambiguously
and will disappear after the integration.

\subsection{Magnetic penguins}

This is another kind penguin correction but with the
magnetic-penguin operator insertion. The corresponding weak
effective Hamiltonian contains the $b\to s g$ transition,
\beq
H_{eff}^{cmp} &=&-\frac{G_F}{\sqrt{2}} V_{tb}V_{ts}^*\; C_{8g}^{eff} O_{8g},
\eeq
with the chromo-magnetic penguin operator,
\beq
O_{8g}&=&\frac{g_s}{8\pi^2}m_b\; \bar{d}_i \; \sigma^{\mu\nu}\; (1+\gamma_5)\;
 T^a_{ij}\; G^a_{\mu\nu}\;  b_j,
\label{eq:o8g}
\eeq
where $i,j$ being the color indices of quarks. The corresponding effective Wilson
coefficient $C_{8g}^{eff}= C_{8g} + C_5$ \cite{nlo05}.

The decay amplitudes obtained by evaluating the Feynman diagrams Fig.\ref{fig:fig4}a
and  Fig.\ref{fig:fig4}b can be written as:
\beq
M^{(g)}_{K\eta_s}&=& \frac{8}{\sqrt{6}} C_F^2 m^6\;
\int_{0}^{1}d x_{1}dx_{2}\,d x_{3}\,\int_{0}^{\infty} b_1d b_1 b_2db_2\,\phi_B(x_1,b_1)
\non
&& \cdot \left\{ \left\{ -\left(1-x_2\right)
\left [ 2\pka(\xeba) + r_K \left(3\pkp(\xeba)-\pkt(\xeba)\right)
\right.\right.\right.\non
&&\left.\left.\left.
+\rk x_2\left(\pkp(\xeba)+\pkt(\xeba)\right)\right ]
\pesa(\xsba)
\right.\right.\non
&&\left.\left.
-\res\left(1+x_2\right) x_3\pka(\xeba)
\left(3 \pesp(\xsba)+\pest(\xsba)\right)
\right.\right.\non
&&\left.\left.
-\rk\res\left(1-x_2\right)\left(\pkp(\xeba)+\pkt(\xeba)\right)\left
(3\pesp(\xsba) -\pest(\xsba)\right)
\right.\right.\non
&&\left.\left.
-\rk\res x_3\left(1-2x_2\right)\left(\pkp(\xeba)-\pkt(\xeba)\right)\left(3\pesp(\xsba)+\pest(\xsba)
\right)\right \} \right.\non
&&\left.
\ \ \cdot E_g(t_q)h_g(A,B,C,b_1,b_2,b_3,x_2)
\right.\non
&&\left.
-\left[4\rk\pkp(\xeba)\pesa(\xsba)+2\rk\res
x_3\pkp(\xeba)\left(3\pesp(\xsba)+\pest(\xsba)\right)\right]
\right.\non
&&\left.
\ \ \cdot E_g(t_q^{\prime})h_g(A^{\prime},B^{\prime},C^{\prime},b_2,b_1,b_3,x_1)
\right\},
\label{eq:mpp1}
\eeq
\beq
M^{(g)}_{\eta_q K} &=& \frac{8}{\sqrt{6}} C_F^2 m^6\;
\int_{0}^{1}d x_{1}dx_{2}\,d x_{3}\,\int_{0}^{\infty} b_1d b_1 b_2db_2\,\phi_B(x_1,b_1)
\non
&& \cdot \left\{ \left\{ -\left(1-x_2\right)
\left [ 2\peqa(\xeba) + \re \left(3\peqp(\xeba)-\peqt(\xeba)\right)
\right.\right.\right.\non
&&\left.\left.\left.
+\re x_2\left(\peqp(\xeba)+\peqt(\xeba)\right)\right ]
\pka(\xsba)
\right.\right.\non
&&\left.\left.
-\rk \left(1+x_2\right) x_3\peqa(\xeba)
\left(3 \pkp(\xsba)+\pkt(\xsba)\right)
\right.\right.\non
&&\left.\left.
-\re \rk \left(1-x_2\right)\left(\peqp(\xeba)+\peqt(\xeba)\right)\left
(3\pkp(\xsba) -\pkt(\xsba)\right)
\right.\right.\non
&&\left.\left.
-\re \rk x_3\left(1-2x_2\right)\left(\peqp(\xeba)-\peqt(\xeba)\right)
\left(3\pkp(\xsba)+\pkt(\xsba)
\right)\right \} \right.\non
&&\left.
\ \ \cdot E_g(t_q)h_g(A,B,C,b_1,b_2,b_3,x_2)
\right.\non
&&\left.
-\left[4\re \peqp(\xeba)\pka(\xsba)+2\re\rk
x_3\peqp(\xeba)\left(3\pkp(\xsba)+\pkt(\xsba)\right)\right]
\right.\non
&&\left.
\ \ \cdot E_g(t_q^{\prime})h_g(A^{\prime},B^{\prime},C^{\prime},b_2,b_1,b_3,x_1)
\right\}.
\label{eq:mpp2}
\eeq
Here $\re =m_0^q /m_B$, $\res=m_0^s/m_B$.
The evolution factors in Eqs.~(\ref{eq:mpp1},\ref{eq:mpp2}) take the form of
\beq
E^{(g)}(t,l^2)&=& \alpha^2_s(t)\; C^{eff}_{8g}(t)\; \exp[-S_{mg}(t)],
\label{eq:egtl2}
\eeq
with the Sudakov factor $S_{mg}$
\beq
S_{mg}(t) &=& s\left(x_1 m_B/\sqrt{2}, b_1\right) +s\left(x_2 m_B/\sqrt{2}, b_2\right)
+s\left((1-x_2) m_B/\sqrt{2}, b_2\right) \non
 && +s\left(x_3 m_B/\sqrt{2}, b_3\right) +s\left((1-x_3) m_B/\sqrt{2}, b_3\right)
\non
&&
-\frac{1}{\beta_1}\left[\ln\frac{\ln(t/\Lambda)}{-\ln(b_1\Lambda)}
+\ln\frac{\ln(t/\Lambda)}{-\ln(b_2\Lambda)}+\ln\frac{\ln(t/\Lambda)}{-\ln(b_3\Lambda)}\right],
\label{smg}.
\eeq

The hard function $h_g$ in the chromo-magnetic penguin amplitude is given by
\beq
h_g(A,B,C,b_1,b_2,b_3,x_i)&=& -S_t(x_i)\, K_0(Bb_1)\, K_0(Cb_3)
\int_{0}^{\pi/2}d\theta \tan\theta\non &&\times
J_0(Ab_1\tan\theta)J_0(Ab_2\tan\theta)J_0(Ab_3\tan\theta)
\eeq
with the index $i=1,2$, the threshold re-summation function $S_t(x_i)$ is given
in Eq.~(\ref{eq:stxi}), and
\beq
A&=&\sqrt{x_2}m_B,\quad B=B^{\prime}=\sqrt{x_1 x_2}m_B,\quad C=i\sqrt{(1-x_2)x_3}m_B,\non
A^{\prime}&=&\sqrt{x_1}m_B,\quad B^{\prime}=B, \quad
C^{\prime}=\sqrt{|x_1-x_3|}m_B.
\eeq
Here  The scale $t_q$, $t_{q}^\prime$, and the gluon invariant mass $l^2$
and $l^{\prime 2}$ have been given in Eqs.~(\ref{eq:tq-tqp}) and (\ref{eq:l2lp2}).

Finally, the total chromo-magnetic penguin
contribution to the considered $B \to K \etap$ ($K=K^0, K^+$)
decays can be written as
\beq
M_{K\eta}^{(cmp)} &=&  <K\eta|{\cal H}_{eff}^{cmp}|B>
= - \frac{G_F}{\sqrt{2}}\;  \lambda_t \; \left [ M^{(g)}_{K\eta_s}\; F_2(\phi)
+ M^{(g)}_{\eta_q K}\; F_1(\phi) \right],\label{eq:mketa-cmp}\\
M_{K\etar}^{(cmp)} &=&  <K\etar|{\cal H}_{eff}^{cmp}|B>
= - \frac{G_F}{\sqrt{2}}\;
\lambda_t \; \left [ M^{(g)}_{K\eta_s}\; F_2^\prime(\phi)
+ M^{(g)}_{\eta_q K}\; F_1^\prime(\phi) \right].
\label{eq:mketap-cmp}
\eeq
The mixing parameters $F_1(\phi)$, $F_1^\prime(\phi)$,
$F_2(\phi)$ and $F_2^\prime(\phi)$ have been defined in Eqs.~(\ref{eq:f1f2phi})
and (\ref{eq:f1f2phip}).

\section{Numerical results and Discussions}\label{sec:n-d}

\subsection{Input parameters}

We use the following input parameters \cite{hfag,pdg2006} in the numerical calculations
\beq
f_B &=& 0.21 {\rm GeV}, \quad f_K = 0.16  {\rm GeV},\quad m_{\eta}=547.5{\rm MeV},\non
m_{\eta^{\prime}}&=&957.8{\rm MeV}, \quad m_K=0.49{\rm GeV}, \quad
m_{0K}=1.7{\rm GeV}, \non
M_B &=& 5.279 {\rm GeV}, m_b = 4.8 {\rm GeV}, \quad M_W = 80.41{\rm GeV},\non
\tau_{B^0} &=& 1.527 {\rm ps}, \quad \tau_{B^+} = 1.643 {\rm ps}.
\label{eq:para}
\eeq

For the CKM quark-mixing matrix, we use the Wolfenstein parametrization as given
in Ref.\cite{hfag,pdg2006}.
\beq
V_{ud}&=&0.9745, \quad V_{us}=\lambda = 0.2200, \quad |V_{ub}|=4.31\times 10^{-3},\non
V_{cd}&=&-0.224, \quad  V_{cd}=0.996, \quad V_{cb}=0.0413, \non
|V_{td}|&=& 7.4 \times 10^{-3}, \quad V_{ts}=-0.042, \quad V_{tb}=0.9991,
\label{eq:vckm}
\eeq
with the CKM angles $\beta=21.6^\circ$, $\gamma =60^\circ \pm 20^\circ $ and
$\alpha=100^\circ \pm 20^\circ $.

\subsection{Branching ratios}

Using the known wave functions and the central values of
relevant input parameters, we find the numerical values of the
corresponding form factors at zero momentum transfer:
 \beq
F^{B\to\etap}_0(q^2=0)&=&0.27^{+0.04}_{-0.03}(\omega_b),\non
F^{B\to K}_0(q^2=0)&=&0.37^{+0.06}_{-0.05}(\omega_b),
\eeq
for $\omega_b=0.40\pm0.04$GeV, which agree well with those obtained in
QCD sum rule calculations.

In the B-rest frame, the branching ratio of a general $B \to PP$ decay can be written as
\beq
Br(B\to M_2 M_3) &=& \tau_B\; \frac{1}{16\pi m_B}\; \chi\;
\left | \calm(B\to M_2 M_3) \right|^2,
\label{eq:br-pp}
\eeq
where $\tau_B$ is the lifetime of the B meson, $\chi\approx 1$ is the phase space
factor and equals to unit when the masses of final state light mesons are neglected.
The total decay amplitude in Eq.~(\ref{eq:br-pp}) is defined as
\beq
\calm(B \to M_2 M_3)= <M_2 M_3| \calh_{eff} +\calh_{eff}^{(ql)} + \calh_{eff}^{(cmp)}|B>.
\eeq

Using  the wave functions and the input parameters as specified in
previous sections, it is straightforward  to calculate the CP-averaged branching
ratios for the  considered four $B \to K \etap$ decays,
which are listed in Table \ref{table1}. For
comparison, we also list the corresponding updated
experimental results \cite{hfag} and numerical results evaluated in
the framework of the QCDF approach \cite{npb675}.

\begin{table}[thb]
\begin{center}
\caption{ The pQCD predictions for the branching ratios (in unit of $10^{-6}$).
The label $\rm{LO_{NLOWC}}$ means the LO results
with the NLO Wilson coefficients, and +VC, +QL, +MP, NLO means the
inclusion  of the vertex corrections, the quark loops, the magnetic
penguin, and all the considered NLO corrections, respectively.}
\label{table1}
\vspace{0.2cm}
\begin{tabular}{l | l| l l l l | l | l |l} \hline \hline
 Mode&  LO & $LO_{{\rm NLOWC}}$ & +VC & +QL & +MP &  NLO & Data & QCDF\\
\hline
$B^+ \to K^+\eta$   &4.7 &4.7 &4.3  &4.9 &3.1 &3.2 &$2.6\pm0.6$ &$1.9 ^{+3.0}_{-1.9}$\\
$B^+ \to K^+\etar$  &30.2&46.8&74.6 &48.1&30.2&51.0&$70.5\pm3.5$&$49.1 ^{+45.2}_{-23.6}$\\
$B^0 \to K^0 \eta$  &3.2 &3.4 & 3.1 &3.8 &2.3 &2.1 &$<2.0$      &$1.1 ^{+2.4}_{-1.5}$\\
$B^0 \to K^0 \etar$ &31.3&46.5&69.7 &48.5&20.7&50.3&$68\pm 4$   &$46.5^{+41.9}_{-22.0}$ \\
 \hline \hline
\end{tabular}
\end{center} \end{table}

It is worth stressing  that the theoretical predictions in the
pQCD approach have relatively large theoretical errors induced by
the still large uncertainties of many input parameters, such as
quark masses($m_{u,d}, m_s$), chiral scales ($m_{0K}, m_0^q, m_0^s$),
Gegenbauer coefficients ($a_i^{(K,\eta)}, \cdots$),
$\omega_b$ and the CKM angles ($\alpha, \gamma$), etc.
The NLO pQCD predictions for the CP-averaged branching ratios with the major theoretical errors
are the following
\beq
Br(\ B^+ \to K^+ \eta) &=& \left
[3.2^{+1.2}_{-0.9}(\omega_b)^{+2.7}_{-1.2}(m_s)^{+1.1}_{-1.0}(a^{\eta}_2)\right
] \times 10^{-6} \label{eq:brp-eta1},\\
Br(\ B^+ \to K^+ \eta^{\prime}) &=& \left
[51.0^{+13.5}_{-8.2}(\omega_b)^{+11.2}_{-6.2}(m_s)^{+4.2}_{-3.5}(a^{\eta}_2)
\right ] \times 10^{-6}, \label{eq:brp-eta2}\\
Br(\ B^0 \to K^0 \eta) &=& \left [2.1
^{+0.8}_{-0.6}(\omega_b)^{+2.3}_{-1.0}(m_s)^{+1.0}_{-0.9}(a^{\eta}_2)
 \right ]\times 10^{-6} \label{eq:br0-eta3} ,\\
Br(\ B^0 \to K^0 \eta^{\prime}) &=& \left [50.3
^{+11.8}_{-8.2}(\omega_b)^{+11.1}_{-6.2}(m_s)^{+4.5}_{-2.7}(a^{\eta}_2)
\right ]\times 10^{-6} \label{eq:br0-eta4},
\eeq
The major errors are induced by the uncertainties of $\omega_b=0.4 \pm 0.04$ GeV,
$m_s=130\pm 30$ MeV and Gegenbauer coefficient
$a^{\eta}_2=0.44\pm 0.22$ (here $a_2^\eta$ denotes $a_2^{\eta_q}$ or $a_2^{\eta_s}$),
respectively.

\begin{figure}[tb]
\begin{center}
\includegraphics[scale=0.25]{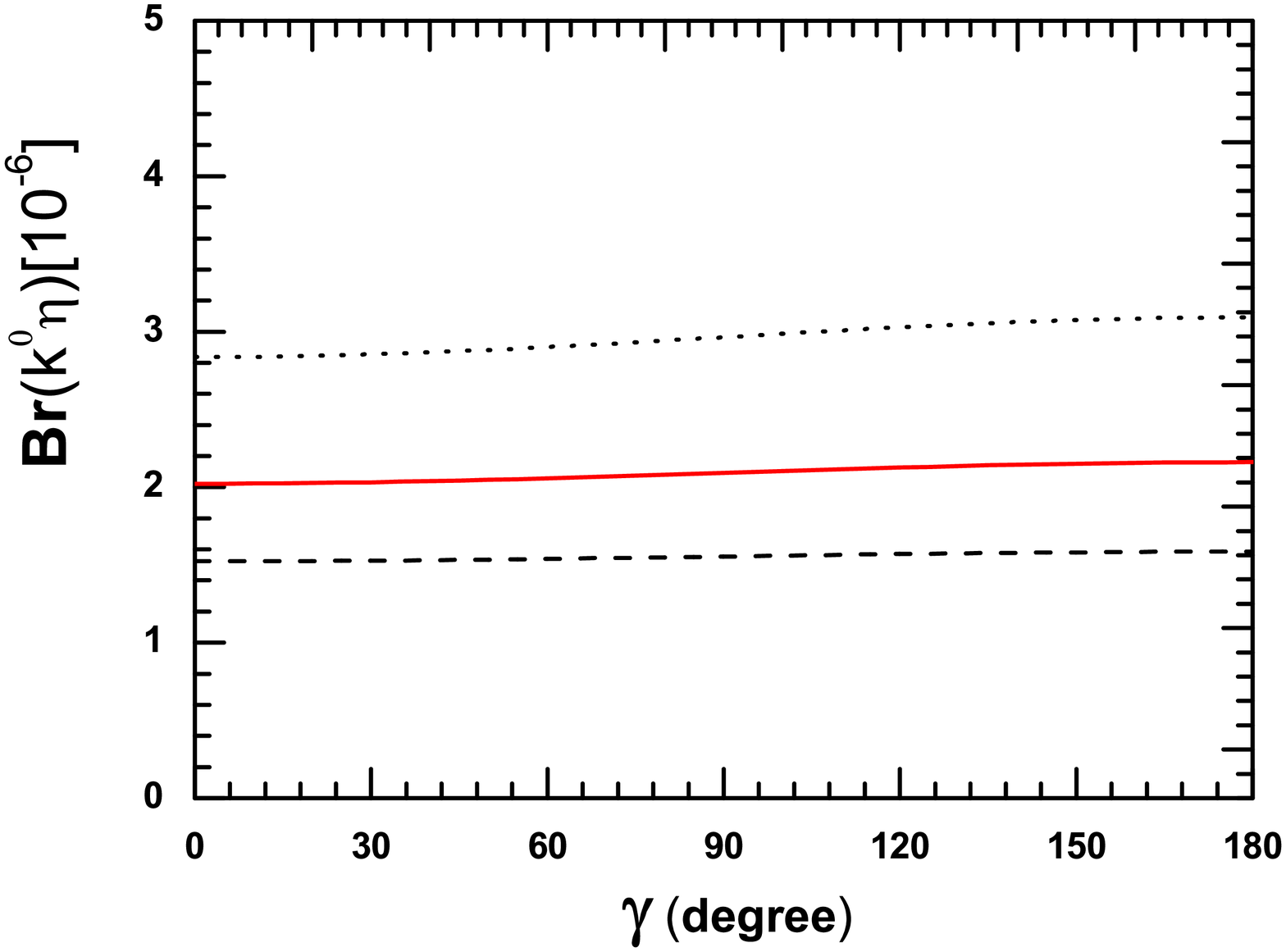}
\includegraphics[scale=0.25]{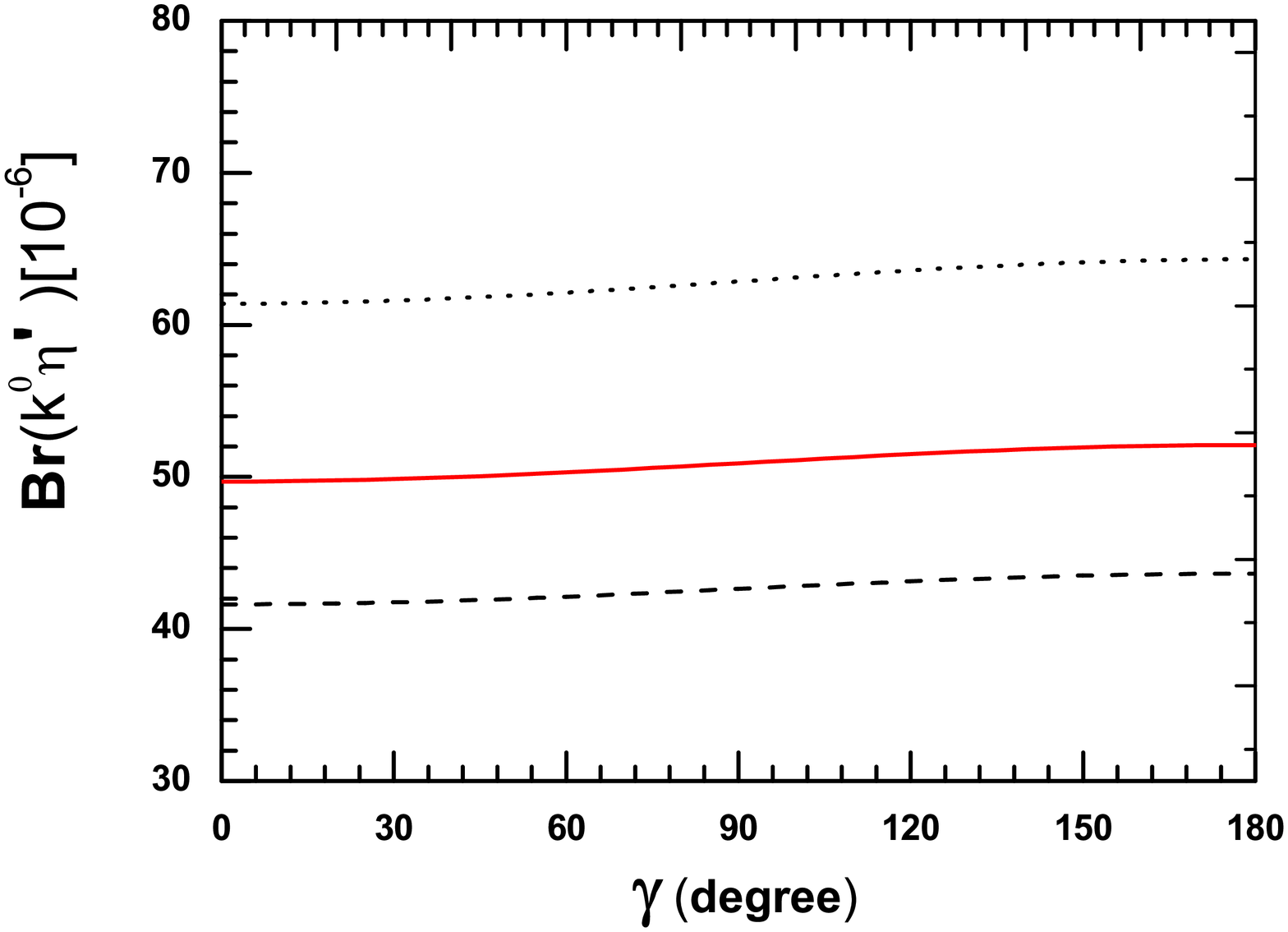}
\vspace{0.3cm}
\caption{The $\gamma$ dependence of the branching
ratios (in units of $10^{-6}$) of $B^0\to K^0\etap$ decays for
$\omega_b=0.36$ GeV (dotted curve), 0.40 GeV (solid curve) and 0.44
GeV (dashed curve). }
\label{fig:fig8}
\end{center}
\end{figure}

\begin{figure}[tb]
\begin{center}
\includegraphics[scale=0.25]{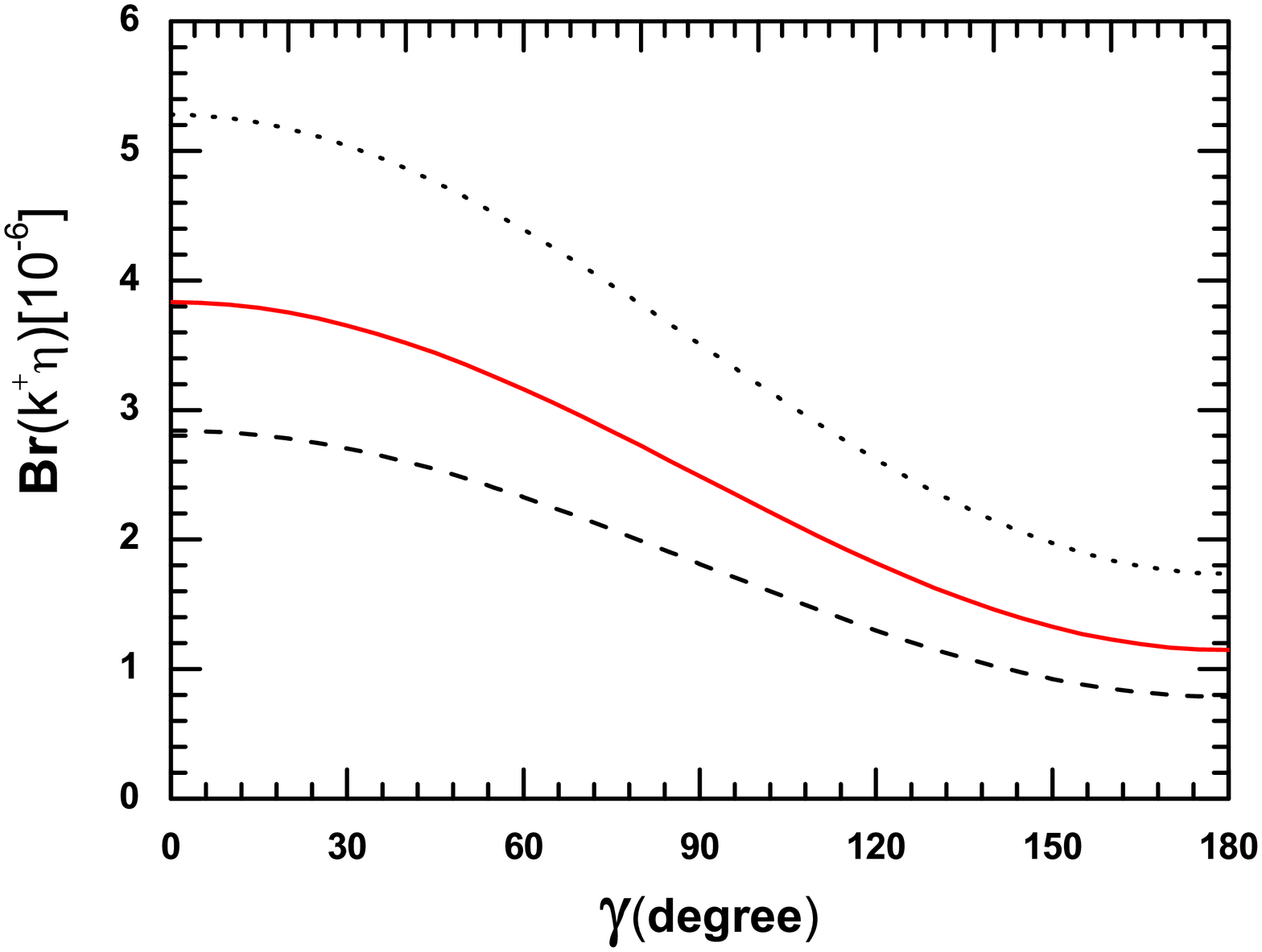}
\includegraphics[scale=0.25]{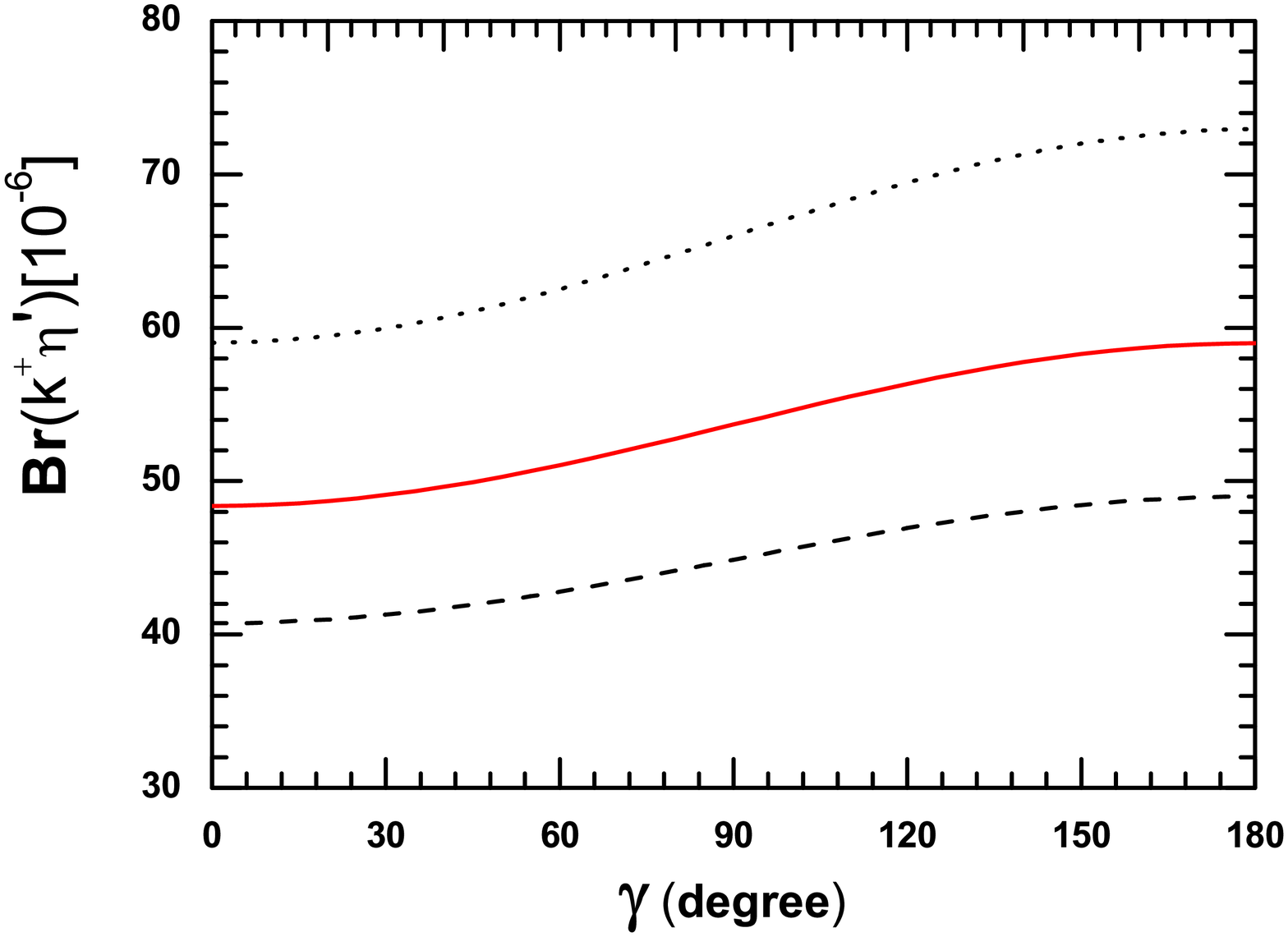}
\vspace{0.3cm}
\caption{The $\gamma$ dependence of the branching
ratios (in units of $10^{-6}$) of $B^+\to K^+\etap$ decays for
$\omega_b=0.36$ GeV (dotted curve), 0.40 GeV (solid curve) and 0.44
GeV (dashed curve). }
\label{fig:fig9}
\end{center}
\end{figure}

In Figs.~\ref{fig:fig8} and \ref{fig:fig9}, we show the
parameter dependence of the pQCD predictions for the branching
ratios of  $B^+\to K^+\etap$ and $B^0\to K^0\etap$ decays for
$\omega_b=0.4\pm0.04$ GeV, $\gamma=[0^\circ,180^\circ]$.

From the numerical results about the branching ratios, one can see that
\begin{itemize}
\item
The decay amplitude $B \to K \eta_q$ and $B \to K \eta_s$ interfere constructively
for $B \to K \etar$ decays, but destructively for $B \to K \eta$ decays.
This mechanism results in a factor of $6-10$ disparity for the branching
ratios of $B \to K^+ \etar$ and  $B \to K^0 \eta$ decays.

\item
The LO pQCD predictions for branching ratios are
much smaller (larger ) than the measured values for $B \to K \etar $ ($B\to K \eta$)
decays, show the same tendency as found in Ref.~\cite{kou02}.

\item
The NLO contributions can interfere constructively (destructively) with the corresponding
LO parst for $B \to K \etar$ ( $B \to K \eta$) decays.
For  $B^0 \to K^0 \etar$ and $B^+ \to K^+ \etar$ decays, the NLO contributions
provide a $70\%$ enhancement to their branching ratios .
For $B^0 \to K^0 \eta$ and $B^+ \to K^+ \eta$ decays, on the other hand,
the NLO contributions give rise to a $30\%$ reduction to their branching ratios
and result in the good agreement between the pQCD predictions and the data.

\item
The NLO pQCD predictions for branching ratios $Br(B \to K \etap)$
agree very well with the measured values within one standard deviation.
The NLO contributions play an important role in
understanding the observed pattern of branching ratios of the
four $B \to K \etap$ decays.

\end{itemize}

\subsection{CP-violating asymmetries}

   Now we turn to the evaluations of the CP-violating asymmetries of $B
\to K \etap$ decays in pQCD approach.
For $B^+ \to K^+ \etap$ decays, the direct CP-violating
asymmetries $\acp$ can be defined as:
 \beq
\acp^{dir} =  \frac{|\overline{\cal M}_f|^2 - |{\cal M}_f|^2}{
 |\overline{\cal M}_f|^2+|{\cal M}_f|^2},
\label{eq:acp1}
\eeq

Using Eq.~(\ref{eq:acp1}), it is easy to calculate the
direct CP-violating asymmetries for the considered
decays, which are listed in Table \ref{table2}. As a comparison, we also list
currently available data \cite{hfag} and the corresponding QCDF predictions \cite{npb675}.

\begin{table}[thb]
\begin{center}
\caption{ The pQCD predictions for the direct CP asymmetries in the NDR scheme
 (in units of $10^{-2}$), the QCDF predictions \cite{npb675}
 and the world average as given by HFAG
 \cite{hfag}.}
\label{table2}
\vspace{0.2cm}
\begin{tabular}{l  |c |c c c c| l |c| c} \hline \hline
 Mode&  LO & $LO_{\rm{NLOWC}}$ & +VC & +QL & +MP &  NLO & Data & QCDF \\
\hline
$\acp^{dir}(B^\pm \to K^\pm \eta)  $ &$9.3  $&$10.3 $&$31.1  $&$7.8  $&$7.6  $&$-11.7 $&$-27\pm 9$  &$-18.9^{+29.0}_{-30.0}$\\
$\acp^{dir}(B^\pm \to K^\pm \etar) $ &$-10.1$&$-7.3 $&$-10.6 $&$-5.9 $&$-10.4$&$-6.2  $&$1.6\pm 1.9$ &$-9.0^{+10.6}_{-16.2}$ \\
 \hline \hline
\end{tabular}\end{center}
\end{table}

The NLO pQCD predictions for $\acp^{dir}(B^+ \to K^+ \etap)$ (in unit of $10^{-2}$)
with the major theoretical errors  are
\beq
\acp^{dir}(B^\pm \to K^\pm \eta) &=& -11.7^{+6.8}_{-9.6}(m_s)^{+3.9}_{-4.2}(\gamma)
^{+2.9}_{-5.6}(a^{\eta_q}_2) \label{eq:acp-a},\quad \non
\acp^{dir}(B^\pm \to K^\pm \etar) &=& -6.2^{+1.2}_{-1.1}(m_s)^{+1.3}_{-1.0}(\gamma)
^{+1.3}_{-1.0}(a^{\eta_q}_2) \label{eq:acp-b},
\eeq
where the dominant errors come from the variations of $m_s=130 \pm 30$ MeV,
$\gamma=60^\circ \pm 20^\circ$ and Gegenbauer coefficient $a^{\eta_q}_2=0.115\pm 0.115$,
respectively.

As to the CP-violating asymmetries for the neutral decays $B^0 \to
K^0 \etap$, the effects of $B^0-\bar{B}^0$ mixing should be considered.
The CP-violating asymmetry of $B^0(\bar B^0) \to K^0 \etap$ decays are time dependent
and can be defined as
\beq
A_{CP} &\equiv& \frac{\Gamma\left
(\overline{B_d^0}(\Delta t) \to f_{CP}\right) -
\Gamma\left(B_d^0(\Delta t) \to f_{CP}\right )}{ \Gamma\left
(\overline{B_d^0}(\Delta t) \to f_{CP}\right ) + \Gamma\left
(B_d^0(\Delta t) \to f_{CP}\right ) }\non
&=& A_{CP}^{dir} \cos(\Delta m  \Delta t) + A_{CP}^{mix} \sin (\Delta m \Delta t),
\label{eq:acp-def}
\eeq
where $\Delta m$ is the mass difference
between the two $B_d^0$ mass eigenstates, $\Delta t
=t_{CP}-t_{tag} $ is the time difference between the tagged $B^0$
($\overline{B}^0$) and the accompanying $\overline{B}^0$ ($B^0$)
with opposite b flavor decaying to the final CP-eigenstate
$f_{CP}$ at the time $t_{CP}$. The direct and mixing induced
CP-violating asymmetries $\acp^{dir}$ ( or ${\cal A}_f$ in term of Belle Collaboration)
and $\acp^{mix}$ can be written as
\beq
\acp^{dir}= {\cal A}_f = \frac{ \left | \lambda_{CP}\right |^2-1 } {1+|\lambda_{CP}|^2},
\qquad \acp^{mix}={\cal S}_f= \frac{ 2 Im (\lambda_{CP})}{1+|\lambda_{CP}|^2},
\label{eq:acp-dm}
\eeq
with the CP-violating parameter $\lambda_{CP}$
\beq
\lambda_{CP} \equiv \left ( \frac{q}{p} \right )_d \cdot
\frac{ \langle f_{CP} |H_{eff}|\overline{B}^0\rangle} {\langle f_{CP} |H_{eff}|B^0\rangle}.
\label{eq:lambda2}
\eeq

By integrating the time variable $t$, one finds the total CP
asymmetries for $B^0 \to K^0 \etap$ decays,
\beq
\acp^{tot}=\frac{1}{1+x^2} A_{CP}^{dir} + \frac{x}{1+x^2} A_{CP}^{mix},
\eeq
where $x=\Delta m/\Gamma=0.775$ \cite{pdg2006}.

In Table \ref{table3}, we show the pQCD predictions for
the central values of the direct, mixing-induced
and total CP asymmetries for $B^0 \to K^0_S \etap $ decays, obtained by
using the LO or NLO Wilson coefficients,
and adding the vertex corrections, the quark loops, the magnetic
penguin, or include all the mentioned NLO corrections, respectively.

\begin{table}[thb]
\begin{center}
\caption{ The pQCD predictions for the direct,mixing induced and total
CP asymmetries (in units of $10^{-2}$) for $B^0 \to K^0 \etap$ decays,
and the world average as given by HFAG \cite{hfag}.}
\label{table3}
\vspace{0.2cm}
\begin{tabular}{l | c |c c c c |c |c } \hline \hline
 Mode&  LO & $LO_{\rm{NLOWC}}$ & +VC & +QL & +MP &  NLO & Data \\\hline
$\acp^{dir}(B^0\to K^0_S \eta) $ &$ -4.2 $&$-1.5$&$-11.2$&$ -0.9 $&$-1.9$&$-12.7$&$--$           \\
$\acp^{dir}(B^0\to K^0_S \etar)$ &$ 1.4  $&$0.0 $&$1.5  $&$ 0.7  $&$-0.1$&$2.3  $&$9\pm 6$ \\ \hline
$\acp^{mix}(B^0\to K^0_S \eta) $ &$ 61.6 $&$67.3$&$ 64.4$&$ 66.9 $&$67.9$&$61.9 $&$--$     \\
$\acp^{mix}(B^0\to K^0_S \etar)$ &$ 64.6 $&$63.5$&$ 63.4$&$ 63.2 $&$63.2$&$62.7 $&$61\pm 7$\\ \hline
$\acp^{tot}(B^0\to K^0_S \eta) $ &$ 27.2 $&$31.7$&$ 24.2$&$ 31.9 $&$31.7$&$22.1 $&$--$     \\
$\acp^{tot}(B^0\to K^0_S \etar)$ &$ 32.1 $&$30.8$&$ 31.7$&$ 31.0 $&$30.5$&$31.8 $&$--$     \\
 \hline \hline
\end{tabular}
\end{center} \end{table}

The NLO pQCD predictions for $\acp^{dir}(B^0 \to K^0 \etap)$ and  $\acp^{mix}(B^0 \to K^0
\etap)$ (in unit of $10^{-2}$) with the major theoretical errors  are
\beq
\acp^{dir}(B^0 \to K^0_S \eta) &=& -12.7\pm4.1(m_s)^{+3.2}_{-1.5}(\gamma)
^{+3.2}_{-6.7}(a^{\eta_q}_2), \label{eq:acp-c} \non
\acp^{dir}(B^0 \to K^0_S \eta^\prime) &=& 2.3 ^{+0.5}_{-0.4}(m_s)^{+0.3}_{-0.6}(\gamma)
^{+0.2}_{-0.1}(a^{\eta_q}_2), \label{eq:acp-d} \non
\acp^{mix}(B^0 \to K^0_S \eta) &=& 61.9 ^{+35.8}_{-65.0}(\gamma)
^{+35.3}_{-64.3}(\alpha), \label{eq:acp-m1}\non
\acp^{mix}(B^0 \to K^0_S \eta^\prime) &=& 62.7^{+35.5}_{-65.0}(\gamma) ^{+35.4}_{-64.7}
(\alpha), \label{eq:acp-m2},
\eeq
where the dominant errors come from the variations of $m_s=130 \pm 30$ MeV,
$\gamma=60^\circ \pm 20^\circ$, $\alpha=100^\circ \pm 20^\circ$,
and the Gegenbauer coefficient $a^{\eta_q}_2 =0.115\pm 0.115$, respectively.

\begin{figure}[tb]
\begin{center}
\includegraphics[scale=0.25]{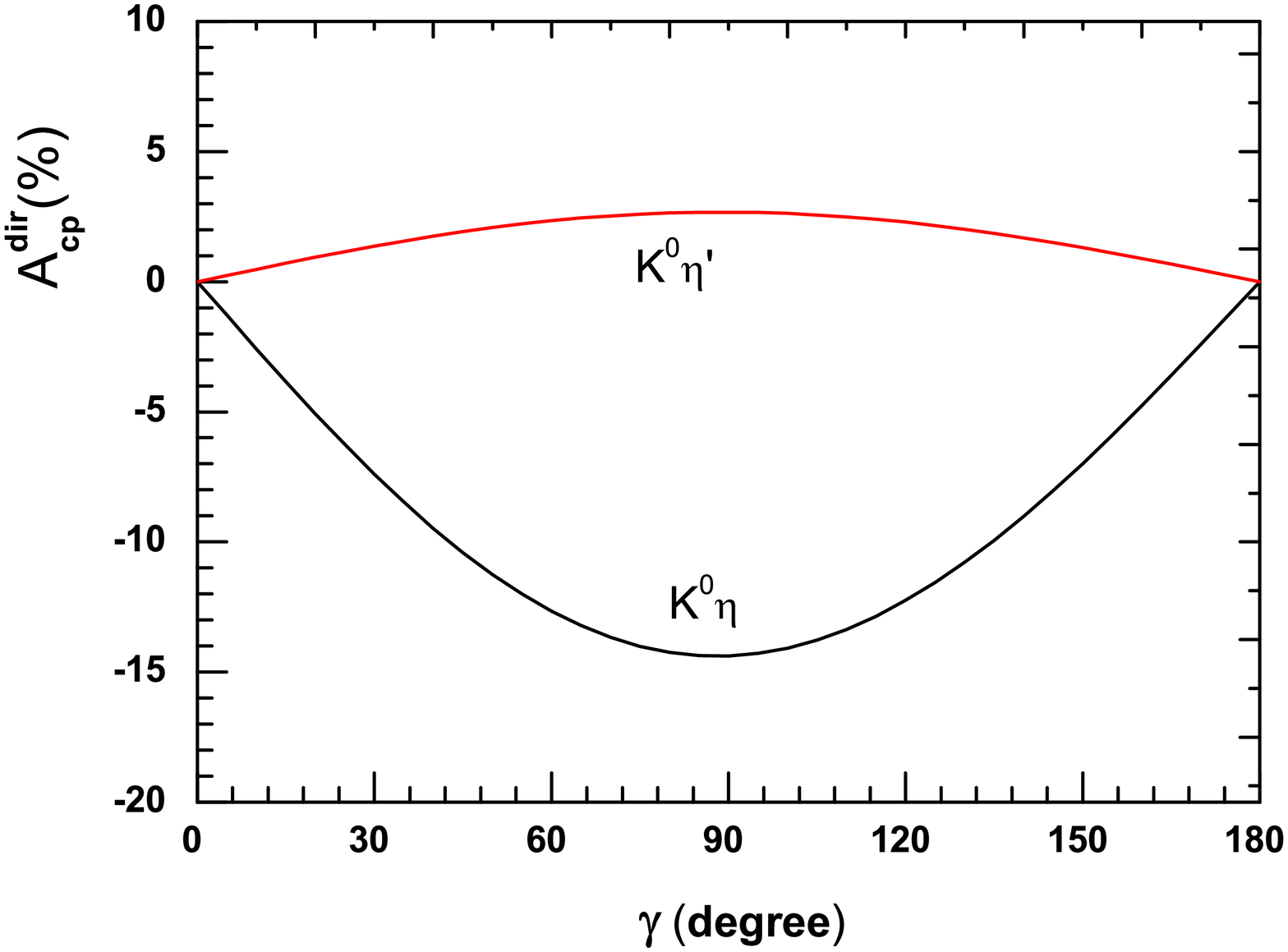}
\includegraphics[scale=0.25]{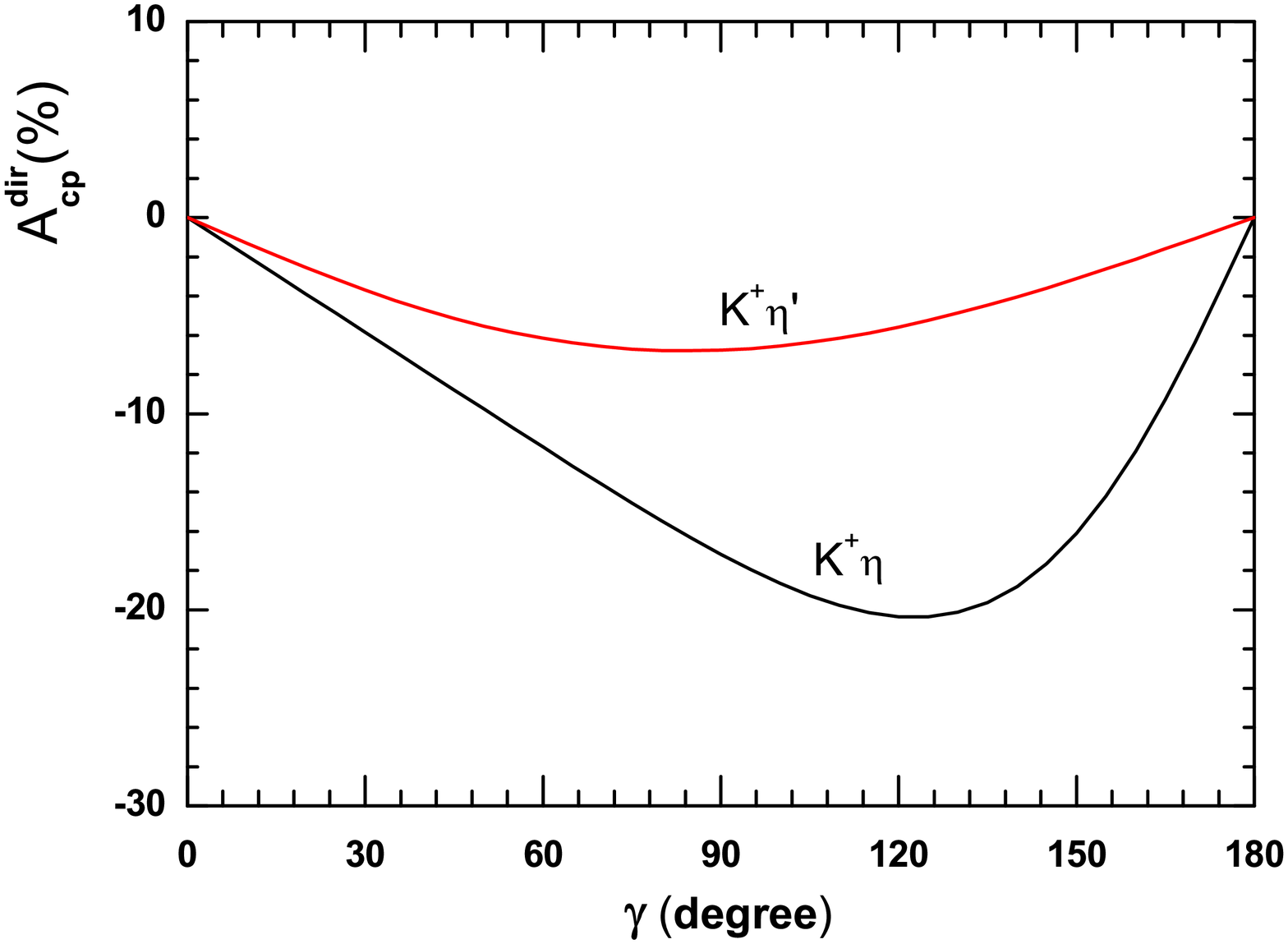}
\vspace{0.3cm}
\caption{The NLO pQCD predictions for direct CP asymmetries (in percentage) of
$B^0 \to K^0_S \etap$ and $B^\pm  \to K^\pm \etap$ decays.}
\label{fig:fig10}
\end{center}
\end{figure}

\begin{figure}[tb]
\begin{center}
\includegraphics[scale=0.26]{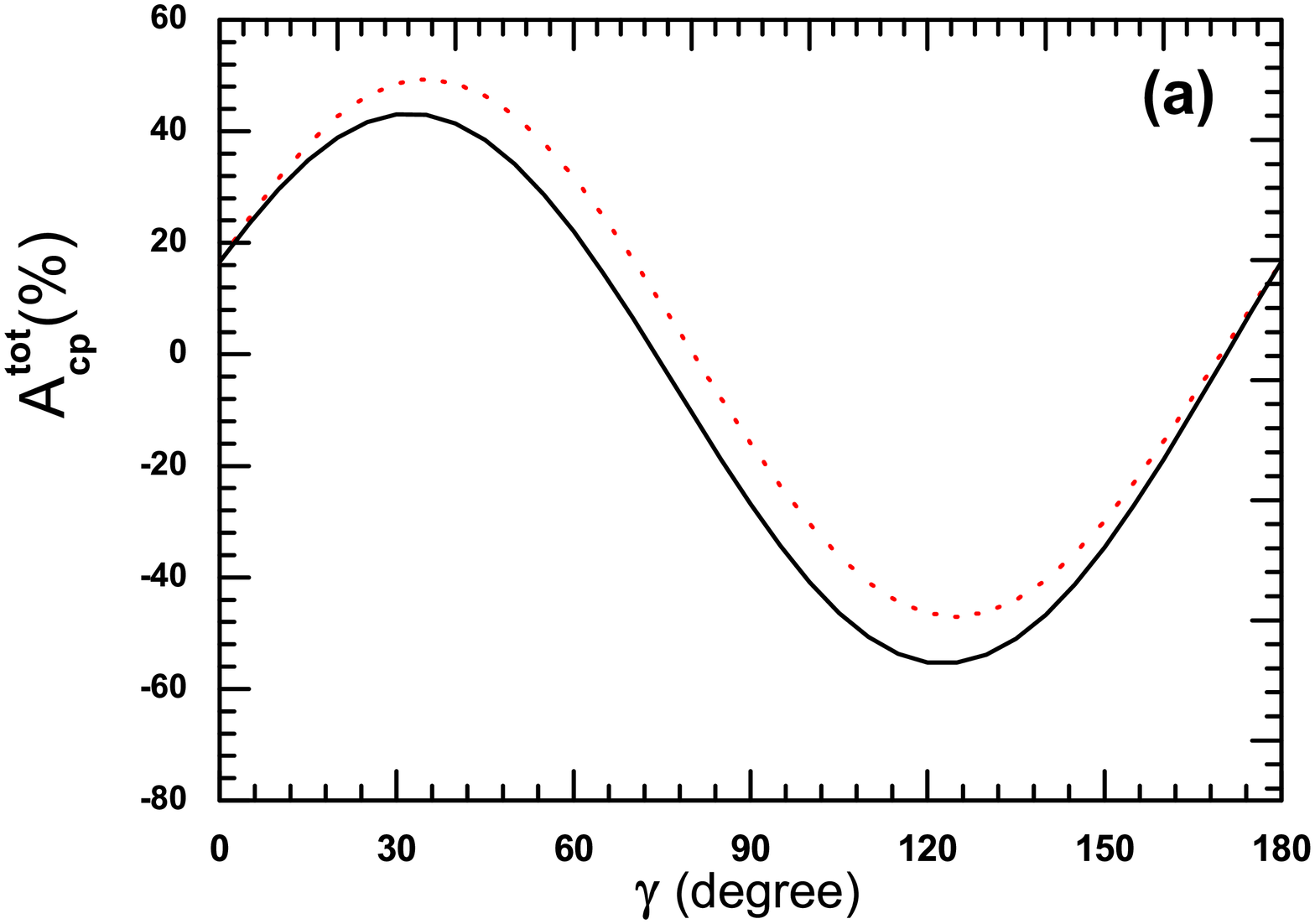}
\includegraphics[scale=0.26]{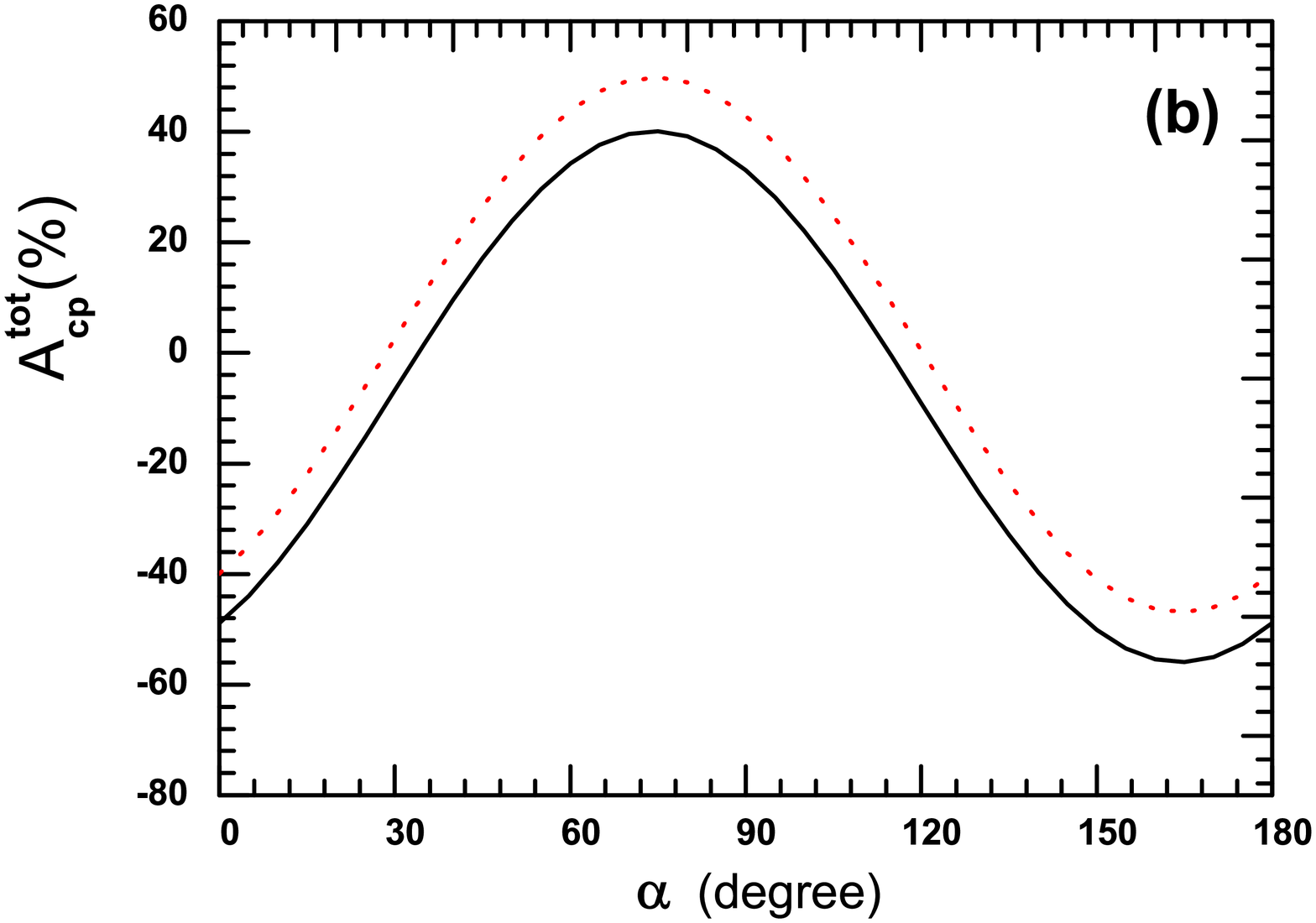}
\vspace{0.3cm}
\caption{The $\gamma$-dependence (a) and the
$\alpha$-dependence (b) of the total CP-asymmetries of $B^0 \to
K^0_S \eta$ (solid curve) and $B^0 \to K^0_S \eta^{\prime}$
(dotted curve) decays. }
\label{fig:fig11}
\end{center}
\end{figure}

In Fig.~\ref{fig:fig10}, we shown the $\gamma$-dependence of the pQCD predictions for
direct CP-violating asymmetries of $B^0 \to K^0_S \etap$ and $B^+ \to K^+ \etap$ decays.
In Fig.~\ref{fig:fig11}, we shown the $\alpha$-dependence of the total
CP-violating asymmetries for $B^0 \to K^0_S \eta$ (solid curve) and $B^0\to
K^0_S\etar$ (dotted curve), respectively.

From the pQCD predictions and currently available experimental measurements for the CP violating
asymmetries of the four $B \to K \etap$ decays, one can see the following points:
\begin{itemize}

\item[]{(a)}
For $B^+ \to K^+ \eta $ decay, the measured direct CP asymmetry
is 3 standard deviation from zero. The LO pQCD prediction changed its sign and
become consistent with the measured one due to the inclusion of NLO
contributions.

\item[]{(b)}
For $\acp^{dir}(B^\pm \to K^\pm \etar)$, the pQCD prediction is changed from $-10\%$
to $-6\%$ due to the inclusion of NLO contributions, which is consistent
with the measured zero result within one standard deviation.

\item[]{(c)}
For $B^0 \to K^0 \etap$ decay, the effects of NLO contributions to their CP asymmetries
are rather small, as can be seen from the numerical results as given
in Table \ref{table3}.

\item[]{(d)}
For neutral $B^0 \to K^0 \etap$ decays, the PQCD predictions are
$\acp^{dir}(B^0\to K^0_S \etar)\approx 2.3\%$ and $\acp^{mix}(B^0\to K^0_S \etar)\approx
63\%$, which agree very well with the data: $(9 \pm 6)\%$ and $(61\pm 7)\%$.
This means that the deviation $\Delta S = -\eta_f S_f - \sin2\beta$ for
$B^0 \to K^0 \etar$ decay is also very small in the pQCD approach.

\end{itemize}

\section{summary }

In this paper,  we calculated the branching ratios and CP-violating
asymmetries of $B^+ \to K^+ \etap$ and $B^0 \to K^0 \etap$
decays in the pQCD approach.
The partial NLO contributions considered here include: QCD vertex corrections,
the quark-loops and the chromo-magnetic penguins.

From our calculations and phenomenological analysis, we found the following results:
\begin{itemize}
\item[]{(a)}
The pQCD predictions for the form factors of $B\to \etap$ and $B \to K$ transitions are
 \beq
F^{B\to\eta}_0(q^2=0)&=&0.21\pm 0.03(\omega_b),\non
F^{B\to\eta^\prime}_0(q^2=0)&=&0.17\pm 0.03 (\omega_b),\non
F^{B\to K}_0(q^2=0)&=&0.37^{+0.06}_{-0.05}(\omega_b),
\eeq
for $\omega_b=0.40\pm0.04$ GeV, which agree well with those obtained in QCD sum rule.

\item[]{(b)}
For branching ratios, the NLO pQCD predictions (in unit of $10^{-6}$) are
\beq
Br(B^+ \to K^+ \eta) &=& 3.2 ^{+3.2}_{-1.8},\non
Br(B^\pm \to K^\pm \etar)&=& 51.0 ^{+18.0}_{-10.9}, \non
Br(B^0 \to K^0 \eta) &=& 2.1 ^{+2.6}_{-1.5},\non
Br(B^0 \to K^0 \etar) &=& 50.3^{+16.8}_{-10.6},
\eeq
where the individual theoretical errors have been added in quadrature.
The decay amplitude $B \to K \eta_q$ and $B \to K \eta_s$ interfere constructively
for $B \to K \etar$ decays, but destructively for $B \to K \eta$ decays.
The NLO contributions in the pQCD approach, furthermore,
can provide a $70\%$ enhancement to $Br(B \to K \etar)$, but
a $30\%$ reduction to $Br(B \to K \eta)$. The large branching ratio of $B \to K \etar$ decays, as
well as the large  disparity  $Br(B \to K \etar) \gg Br(B \to K \eta)$ can therefore be
understood naturally.

\item[]{(c)}
The pQCD predictions for the CP asymmetries of $B \to K \etap$ decays are consistent
with currently available data. For neutral $B^0 \to K^0 \etap$ decays, for example,
the PQCD predictions are $\acp^{dir}(B^0\to K^0_S \etar)\approx 2.3\%$ and
$\acp^{mix}(B^0\to K^0_S \etar)\approx 63\%$, which  agree very well with the measured values
of $(9 \pm 6)\%$ and $(61\pm 7)\%$, respectively.

\item[]{(d)}
In this paper, only the partial NLO contributions have been taken into account.
We think that they are the dominant part of the whole NLO corrections.
To achieve a complete NLO calculations in the pQCD approach, the still missing
pieces from the emission diagrams, hard-spectator and annihilation diagrams,
should be evaluated as soon as possible.

\end{itemize}

\begin{acknowledgments}

The authors are very grateful  to Hsiang-nan Li, Cai-Dian L\"u, Ying Li, Wei Wang
and Yu-Ming Wang for helpful discussions.
This work is partly supported  by the National Natural Science
Foundation of China under Grant No.10575052 and 10735080.

\end{acknowledgments}


\begin{appendix}

\section{Related Functions }\label{sec:aa}

We show here the function $h_i$'s, coming from the Fourier
transformations  of the hard kernel $H^{(0)}(x_i,b_i)$,
\beq
h_e(x_1,x_2,b_1,b_2)&=&  K_{0}\left(\sqrt{x_1 x_2} m_B b_1\right)
\left[\theta(b_1-b_2)K_0\left(\sqrt{x_2} m_B
b_1\right)I_0\left(\sqrt{x_2} m_B b_2\right)\right. \non & &\;\left.
+\theta(b_2-b_1)K_0\left(\sqrt{x_2}  m_B b_2\right)
I_0\left(\sqrt{x_2}  m_B b_1\right)\right] S_t(x_2), \label{he1}
\eeq \beq h_a(x_2,x_3,b_2,b_3)&=& K_{0}\left(i\sqrt{(1-x_2)x_3} m_B
b_2\right)
 \left[\theta(b_3-b_2)K_0\left(i \sqrt{x_3} m_B
b_3\right)I_0\left(i \sqrt{x_3} m_B b_2\right)\right. \non &
&\;\;\;\;\left. +\theta(b_2-b_3)K_0\left(i \sqrt{x_3}  m_B
b_2\right) I_0\left(i \sqrt{x_3}  m_B b_3\right)\right] S_t(x_3),
\label{eq:he3}
\eeq
\beq
 h_{f}(x_1,x_2,x_3,b_1,b_3) &=& \biggl\{\theta(b_1-b_3)\mathrm{K}_0(M_B\sqrt{x_1 x_2} b_1)
  \mathrm{I}_0(M_B\sqrt{x_1 x_2} b_3)\non &+ & \theta(b_3-b_1)\mathrm{K}_0(M_B\sqrt{x_1 x_2} b_3)
  \mathrm{I}_0(M_B\sqrt{x_1 x_2} b_1) \biggr\} \non && \cdot\left(
\begin{matrix}
 \frac{\pi i}{2}\mathrm{H}_0(\sqrt{(x_2(x_3-x_1))} M_B b_3), & \text{for}\quad x_1-x_3<0 \\
 \mathrm{K}_0^{(1)}(\sqrt{(x_2(x_1-x_3)}M_B b_3), &
 \text{for} \quad x_1-x_3>0
\end{matrix}\right),
\label{eq:pp1} \eeq
\beq h_f^3(x_1,x_2,x_3,b_1,b_3) &=&
\biggl\{\theta(b_1-b_3) \mathrm{K}_0(i \sqrt{(1-x_2) x_3} b_1 M_B)
 \mathrm{I}_0(i \sqrt{(1-x_2) x_3} b_3 M_B)\non &+&(\theta(b_3-b_1) \mathrm{K}_0(i \sqrt{(1-x_2) x_3} b_3 M_B)
 \mathrm{I}_0(i \sqrt{(1-x_2) x_3} b_1 M_B) \biggr\}
 \non
& & \cdot\left(
\begin{matrix}
 \mathrm{K}_0(M_B\sqrt{(x_1-x_3)(1-x_2)}b_1), & \text{for}\quad x_1-x_3>0 \\
 \frac{\pi i}{2} \mathrm{H}_0^{(1)}(M_B\sqrt{(x_3-x_1)(1-x_2)}b_1), &
 \text{for} \quad x_1-x_3<0
\end{matrix} \right),
\label{eq:pp4}
\eeq
 \beq
 h_f^4(x_1,x_2,x_3,b_1,b_2) &=&
 \biggl\{\theta(b_1-b_3) \mathrm{K}_0(i \sqrt{(1-x_2) x_3} b_1 M_B)
 \mathrm{I}_0(i \sqrt{(1-x_2) x_3} b_3 M_B)
 \non
&+& \theta(b_3-b_1) \mathrm{K}_0(i \sqrt{(1-x_2) x_3} b_3
M_B)\mathrm{I}_0(i \sqrt{(1-x_2) x_3} b_1 M_B)\biggr\} \non &&\cdot
\left(
\begin{matrix}
 \mathrm{K}_0(M_B F_{1}b_1), & \text{for}\quad F_{1}^2>0 \\
 \frac{\pi i}{2} \mathrm{H}_0^{(1)}(M_B \sqrt{|F_{1}^2|}b_1), &
 \text{for}\quad F_{1}^2<0
\end{matrix}\right), \label{eq:pp3}
\eeq
where $J_0$ is the Bessel function and $K_0$, $I_0$ are
modified Bessel functions with $K_0 (-i x) = -(\pi/2) Y_0 (x) + i (\pi/2)
J_0 (x)$, and $F_{(1)}$'s are defined by
\beq
F^2_{(1)}&=&1-x_2(1-x_1-x_3).
 \eeq

The threshold resummation form factor $S_t(x_i)$ is adopted
from Ref.~\cite{kurimoto}. It has been parameterized as
\beq
S_t(x)=\frac{2^{1+2c} \Gamma (3/2+c)}{\sqrt{\pi}
\Gamma(1+c)}[x(1-x)]^c,
\label{eq:stxi}
\eeq
where the parameter $c=0.3$. This function is normalized to unity.

The evolution factors $E^{(\prime)}_e$, and $E^{(\prime)}_a$,
appeared in the decay amplitudes are given by
\beq
E_e(t)&=&\alpha_s(t)\exp[-S_{ab}(t)],\non
E_e^{\prime}(t)&=&\alpha_s(t)\exp[-S_{cd}(t)]|_{b_2=b_1},\non
E_a^{\prime}(t)&=&\alpha_s(t)\exp[-S_{ef}(t)]|_{b_2=b_3},\non
E_a(t)&=&\alpha_s(t)\exp[-S_{gh}(t)].
\label{eq:eet01}
\eeq

The Sudakov factors used in the text are defined as
\beq
S_{ab}(t)
&=& s\left(x_1 m_B/\sqrt{2}, b_1\right) +s\left(x_2 m_B/\sqrt{2},
b_2\right) +s\left((1-x_2) m_B/\sqrt{2}, b_2\right) \non
&&-\frac{1}{\beta_1}\left[\ln\frac{\ln(t/\Lambda)}{-\ln(b_1\Lambda)}
+\ln\frac{\ln(t/\Lambda)}{-\ln(b_2\Lambda)}\right],
\label{eq:sab}\\
 S_{cd}(t) &=& s\left(x_1 m_B/\sqrt{2}, b_1\right)
 +s\left(x_2 m_B/\sqrt{2}, b_1\right)
+s\left((1-x_2) m_B/\sqrt{2}, b_1\right) \non
 && +s\left(x_3
m_B/\sqrt{2}, b_3\right) +s\left((1-x_3) m_B/\sqrt{2}, b_3\right)
\non
 & &-\frac{1}{\beta_1}\left[2
\ln\frac{\ln(t/\Lambda)}{-\ln(b_1\Lambda)}
+\ln\frac{\ln(t/\Lambda)}{-\ln(b_3\Lambda)}\right],
\label{scd}\\
S_{ef}(t) &=& s\left(x_1 m_B/\sqrt{2}, b_1\right)
 +s\left(x_2 m_B/\sqrt{2}, b_3\right)
+s\left((1-x_2) m_B/\sqrt{2}, b_3\right) \non
 && +s\left(x_3
m_B/\sqrt{2}, b_3\right) +s\left((1-x_3) m_B/\sqrt{2}, b_3\right)
\non
 &
&-\frac{1}{\beta_1}\left[\ln\frac{\ln(t/\Lambda)}{-\ln(b_1\Lambda)}
+2\ln\frac{\ln(t/\Lambda)}{-\ln(b_2\Lambda)}\right],
\label{sef}\\
S_{gh}(t) &=& s\left(x_2 m_B/\sqrt{2}, b_2\right)
 +s\left(x_3 m_B/\sqrt{2}, b_3\right)
+s\left((1-x_2) m_B/\sqrt{2}, b_2\right) \non
 &+& s\left((1-x_3)
m_B/\sqrt{2}, b_3\right)
-\frac{1}{\beta_1}\left[\ln\frac{\ln(t/\Lambda)}{-\ln(b_3\Lambda)}
+\ln\frac{\ln(t/\Lambda)}{-\ln(b_2\Lambda)}\right],
\label{sgh}
\eeq
where the function $s(q,b)$ are defined in the
Appendix A of Ref.\cite{luy01}. The scale $t_i$'s in the above
equations are chosen as \beq
t_{a} &=& {\rm max}(\sqrt{x_2} m_B,\sqrt{x_1 x_2}m_B,1/b_1,1/b_2)\;,\non
t_{a}^{\prime} &=& {\rm max}(\sqrt{x_1}m_B,\sqrt{x_1 x_2}m_B,1/b_1,1/b_2)\;,\non
t_{b} &=& {\rm max}(\sqrt{x_2|1-x_3-x_1|}m_B,\sqrt{x_1 x_2}m_B,1/b_1,1/b_3)\;,\non
t_{b}^{\prime} &=& {\rm max}(\sqrt{x_2|x_3-x_1|}m_B,\sqrt{x_1 x_2}m_B,1/b_1,1/b_3)\;,\non
t_{c} &=& {\rm max}(\sqrt{(1-x_2) x_3}m_B, \sqrt{|x_1-x_3|(1-x_2)}m_B,1/b_1,1/b_3)\;,\non
t_{c}^{\prime} &=& {\rm max}(\sqrt{|1-x_2(1-x_3-x_1)|}m_B,
    \sqrt{(1-x_2) x_3} m_B,1/b_1,1/b_3)\;,\non
t_{d} &=&{\rm
max}(\sqrt{(1-x_2)x_3}m_B,\sqrt{(1-x_2)}m_B,1/b_2,1/b_3)\;,\non
t_{d}^{\prime} &=&{\rm max}(\sqrt{(1-x_2)x_3} m_B,\sqrt{x_3}m_B,1/b_2,1/b_3)\; .
\eeq
\end{appendix}


\end{document}